\documentclass[journal]{IEEEtran}
\ifCLASSINFOpdf
   \usepackage[pdftex]{graphicx}
\else
   \usepackage[dvips]{graphicx}
\fi
\usepackage[inline]{enumitem}

\usepackage{amsmath}
\usepackage{algorithmic}

\usepackage{array}

\ifCLASSOPTIONcompsoc
 \usepackage[caption=false,font=normalsize,labelfont=sf,textfont=sf]{subfig}
\else
 \usepackage[caption=false,font=footnotesize]{subfig}
\fi

\makeatletter
\long\def\@makecaption#1#2{\ifx\@captype\@IEEEtablestring%
\footnotesize\begin{center}{\normalfont\footnotesize #1}\\
{\normalfont\footnotesize\scshape #2}\end{center}%
\@IEEEtablecaptionsepspace
\else
\@IEEEfigurecaptionsepspace
\setbox\@tempboxa\hbox{\normalfont\footnotesize {#1.}~~ #2}%
\ifdim \wd\@tempboxa >\hsize%
\setbox\@tempboxa\hbox{\normalfont\footnotesize {#1.}~~ }%
\parbox[t]{\hsize}{\normalfont\footnotesize \noindent\unhbox\@tempboxa#2}%
\else
\hbox to\hsize{\normalfont\footnotesize\hfil\box\@tempboxa\hfil}\fi\fi}
\makeatother
\usepackage{booktabs}
\usepackage{stfloats}
\usepackage{pifont}
\newcommand{\cmark}{\ding{51}}%
\newcommand{\xmark}{\ding{55}}%

\usepackage{multirow}
\usepackage{amsfonts}

\usepackage{cleveref}

\begin{document}
%
\title{Voice activity detection in the wild: A data-driven approach using teacher-student training}
%
%

\author{Heinrich Dinkel,~\IEEEmembership{Student Member,~IEEE,}
        Shuai~Wang,~\IEEEmembership{Student Member,~IEEE,},
        Xuenan Xu,~\IEEEmembership{Student Member,~IEEE,}
        Mengyue~Wu$\dagger$,~\IEEEmembership{Member,~IEEE}
        Kai Yu$\dagger$,~\IEEEmembership{Senior Member,~IEEE\thanks{$\dagger$Mengyue Wu and Kai Yu are the corresponding authors.}}
\thanks{}}

%
%

\markboth{Journal of \LaTeX\ Class Files,~Vol.~14, No.~8, August~2015}%
{Shell \MakeLowercase{\textit{et al.}}: Bare Demo of IEEEtran.cls for IEEE Journals}
%



\maketitle

\begin{abstract}
Voice activity detection is an essential pre-processing component for speech-related tasks such as automatic speech recognition (ASR). 
Traditional supervised VAD systems obtain frame-level labels from an ASR pipeline by using, e.g., a Hidden Markov model.
These ASR models are commonly trained on clean and fully transcribed data, limiting VAD systems to be trained on clean or synthetically noised datasets.
Therefore, a major challenge for supervised VAD systems is their generalization towards noisy, real-world data.
This work proposes a data-driven teacher-student approach for VAD, which utilizes vast and unconstrained audio data for training.
Unlike previous approaches, only weak labels during teacher training are required, enabling the utilization of any real-world, potentially noisy dataset.
Our approach firstly trains a teacher model on a source dataset (Audioset) using clip-level supervision.
After training, the teacher provides frame-level guidance to a student model on an unlabeled, target dataset.
A multitude of student models trained on mid- to large-sized datasets are investigated (Audioset, Voxceleb, NIST SRE).
Our approach is then respectively evaluated on clean, artificially noised, and real-world data.
We observe significant performance gains in artificially noised and real-world scenarios.
Lastly, we compare our approach against other unsupervised and supervised VAD methods, demonstrating our method's superiority.
\end{abstract}


%
\IEEEpeerreviewmaketitle

\section{Introduction}

\IEEEPARstart{V}{oice} activity detection (VAD, or speech activity detection, SAD) in some literature, whose main objective is to detect voiced speech segments and distinguish them from unvoiced ones, is crucial as a pre-processing step for tasks such as speech recognition and speaker recognition.

VAD can be performed via either unsupervised feature-based or supervised model-based approaches. 
For feature-based VAD, simple features such as energy~\cite{woo2000robust,povey2011kaldi} and zero-crossing rate~\cite{rabiner1975algorithm,junqua1991study} and more complex ones such as the spectral shape~\cite{rabiner1977application} and pitch~\cite{morales2011pitch,Tan2020} are investigated.
The latter requires speech and non-speech labels for the training data to build statistical models that discriminate between speech or non-speech signals~\cite{sohn1999statistical}. 
Contrary to supervised frameworks, unsupervised methods do not require extensive amounts of labeled data. 
Therefore unsupervised approaches are cheaper to train and often faster (due to simpler architecture) than their supervised counterparts. 
Unsupervised methods are thus a popular research direction in VAD~\cite{Sharma2019,Sadjadi2013UnsupervisedSA,Ying2011VoiceAD,Tao_2016,Zhang2013a,Zhang2013b}.

However, despite the simplicity of unsupervised methods, they suffer from not scaling well with large amounts of data.
On the other hand, supervised model-based VAD can obtain better performance when training data size scales up due to a more accurate estimation of the model parameters.

The choice of backbone models is essential for supervised VAD approaches. 
Before the era of deep learning, statistical models such as the Gaussian mixture model (GMM)~\cite{fukuda2010long} and Hidden Markov model (HMM)~\cite{sohn1999statistical,varela2011combining} are used to model the distribution of speech and non-speech signals.
Deep learning techniques have contributed to the recent success in VAD~\cite{Hughes2013,ryant2013speech,thomas2014analyzing,Kim2018,Lavechin2020,Lee2020}. 
Deep neural networks (DNN)~\cite{Segbroeck2013} and specifically convolutional neural networks (CNN)~\cite{Lin2019,Vafeiadis2019} offer improved modeling capabilities compared to traditional methods~\cite{ryant2013speech}, while recurrent- (RNN) and long short-term memory (LSTM) networks can better model long-term dependencies between sequential inputs~\cite{Hughes2013,eyben2013real,Tong2016,Kim2018}.
Lastly, semi-supervised VAD, which incorporates labeled and unlabeled data, has also been investigated in~\cite{SHOLOKHOV2018132}.
However, despite the recent success of deep learning models in VAD, supervised frame-level labels are required for training. 

Most methods currently acquire those labels via an automatic speech recognition (ASR) pipeline, where frame-level speech activation is estimated via an HMM model trained on transcribed, clean data.
Accordingly, the prerequisite includes both prior knowledge about the spoken language (phonemes) and clean training data, and therefore, such methods cannot easily scale with arbitrary data.
Thus training data is usually recorded under a controlled environment with or without additional synthetic noise~\cite{hirsch2000aurora,Tong2016}, with work aiming at de-noising~\cite{Jung2018,Zhang2013b,Ghosh2018}.
However, only having access to synthetic noise inevitably prevents VAD from generalizing to real-world applications, where speech in the wild is often accompanied by countless unseen sounds, each with its unique features. 
Moreover, real-world data is likely to contain copious amounts of spoken language data mixed with any arbitrary noise, challenging to be used in traditional supervised VAD frameworks.

Recent work in~\cite{Dinkel2020a} proposed general-purpose VAD (GPVAD), a framework using weak labeled supervision (on clip-level), as an alternative to common supervised VAD approaches.
However, while the proposed GPVAD framework in~\cite{Dinkel2020a} outperforms strongly supervised VAD when evaluating on real-world data, GPVAD's clean and synthetic noise performance is inferior to traditional supervised VAD approaches.
We believe the inferior GPVAD performance stems mainly from two factors:
\begin{enumerate}
    \item Strongly supervised VAD models have access to frame-level labels, enhancing their capability to estimate speech duration.
    \item Language/Phonetic unit match between training and evaluation datasets (e.g., English).
\end{enumerate}

One possible advantage of GPVAD against traditional supervised VAD methods is that data collection is comparatively cheap since real-world publicly available datasets can be used, and only clip-level labels are required.
This work aims to address the two problems stated above by extending the GPVAD framework towards a generalized data setting.
We adopt a teacher-student approach and estimate frame-level labels for the student model from weakly-labeled teacher training.
Therefore, this study aims to provide insight if VAD models can improve noise robustness by utilizing large amounts of data without requiring manual frame-level annotation or exclusively rely on clean data. 

The paper is organized as follows: In \Cref{sec:approach}, we introduce our method. 
Further, in \Cref{sec:experiments}, the experimental setup and training details and evaluation schemes are provided. 
Then, in \Cref{sec:vad_results}, our results are provided and analyzed regarding their noise robustness in VAD. 
Finally, a summary is provided in \Cref{sec:conclusion}.

\section{VAD in the wild}
\label{sec:approach}

\begin{figure*}
    \centering
    \includegraphics[width=\linewidth]{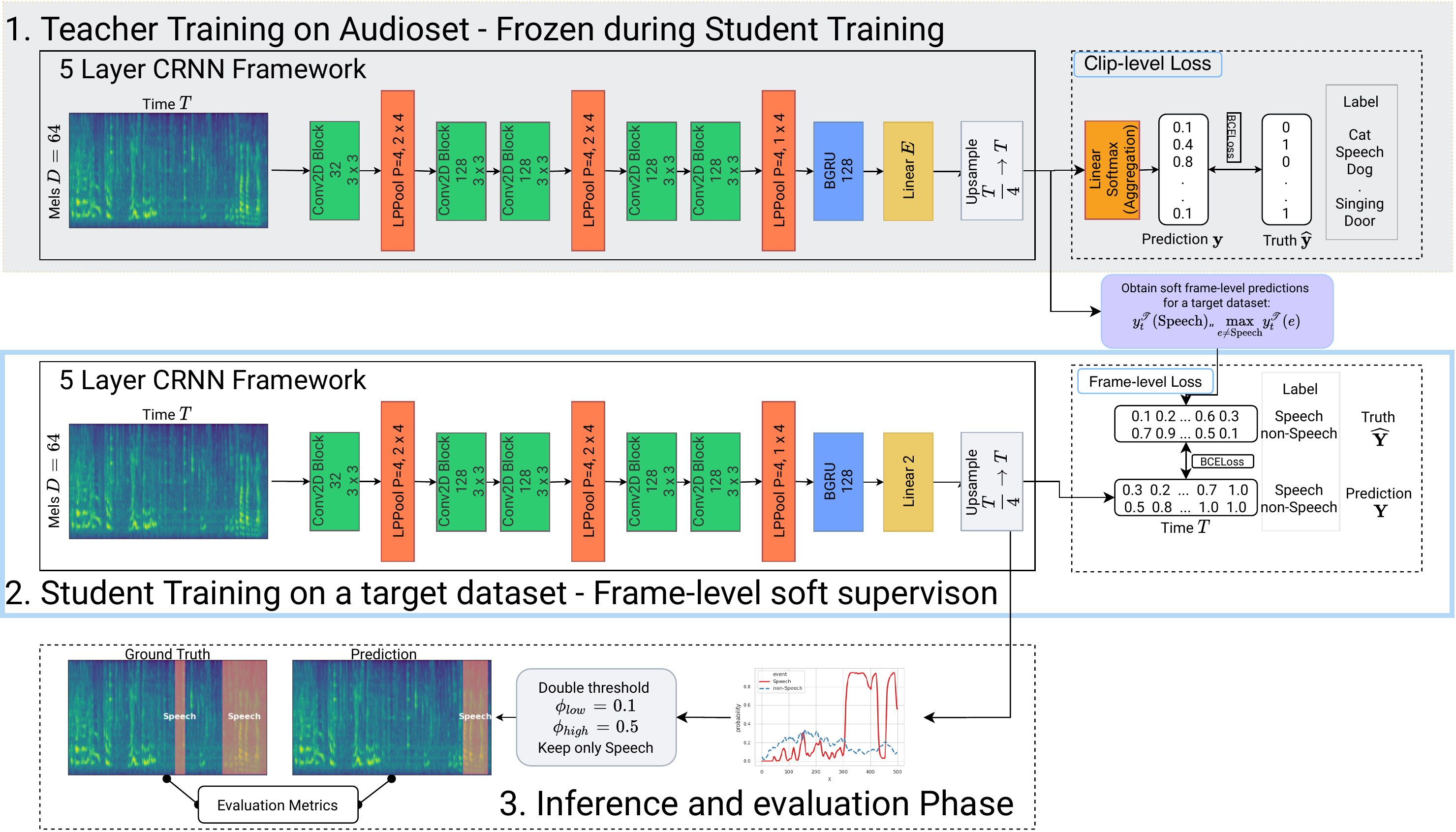}
    \caption{The proposed data-driven VAD framework. A convolution block refers to an initial batch normalization, then a $3\times3$ convolution, and lastly, a LeakyReLU (slope $-0.1$) activation. All convolutions use padding to preserve the input size. The framework consists of three distinct stages: 1. Clip-level training of a teacher model on source data (Audioset). 2. Using the teacher to estimate soft labels for a student model on a target dataset. 3. Evaluation of the student model by only keeping the Speech class.}
    \label{fig:data_driven_vad}
\end{figure*}

Traditionally, VAD for noisy scenarios is modeled as in \Cref{eq:noisy_eq}.
The assumption is that additive noise $\mathbf{u}$ can be filtered out from an observed speech signal $\mathbf{x}$ to obtain clean speech $\mathbf{s}$.
\begin{equation}
\label{eq:noisy_eq}
    \mathbf{x} = \mathbf{s} + \mathbf{u}
\end{equation}
Conventional approaches tackle the problem from a signal processing perspective, where the noised signal $\mathbf{x}$ is filtered by a multitude of low- and high-pass filters, as well as other noise suppression techniques to remove $\mathbf{u}$~\cite{povey2011kaldi,Tan2020,Tan2010}. 
However, VAD systems trained with this framework cannot scale easily with real-world data since directly modeling $\mathbf{u}$ with various noise types is difficult.
Therefore, we aim at learning the properties of $\mathbf{s}$ accompanied with potentially $L$ different non-speech events $\mathbf{U} = \left( \mathbf{u}_0, \mathbf{u}_1,\ldots,\mathbf{u}_L \right)$, where $\mathbf{u}_0 = 0$.

\begin{align}
\label{eq:model}
    \begin{split}
        \mathcal{X} &= \{ \mathbf{x}_1,\ldots, \mathbf{x}_{l}, \ldots , \mathbf{x}_{L} \}\\
        \mathbf{x}_l &= \left( \mathbf{s}, \mathbf{u}_l \right)
    \end{split}
\end{align}

Here, we model our observed speech data $\mathcal{X}$ as a ``bag'', containing all co-occurrences of \textit{Speech} in conjunction with another, possibly noisy background/foreground event label $l \in \{0, \ldots, L\}$ from a set of all possible event labels $L < E$ (\Cref{eq:model}).
Here $E$ is the total number of event labels observed.
Since our approach stems from weakly supervised sound event detection (WSSED), we do not restrict our approach to only model $L$ event, and instead, we aim at modeling all $E$ events.
This potentially enhances our model's robustness since it not only has access to speech-only data, commonly seen in traditional VAD approaches but also to data in the wild.

\subsection{Teacher-student approach}

This work proposes a data-driven teacher-student VAD approach, which only requires weak clip labels during training.
The approach is based on WSSED, which detects and localizes different sounds, including speech, via clip-level supervision.
Specifically, the approach estimates from a given input audio-clip spectrogram $\mathbf{S} \in \mathbb{R}^{T\times D}$ with duration $T$ (here number of frames) and $D$ frequency-bins, a clip-level label $y$ as:
\begin{align}
\begin{split}
\left[y_1, \ldots, y_T\right] = F\left(\mathbf{S}\right)\\
  y = \Gamma\left[y_1, \ldots, y_T\right]
\end{split}
\end{align}

, where $F$ is modeled via a neural network.
Note that the temporal pooling function $\Gamma$, which removes all time-variability, is the only direct connection between the observed, weakly supervised signal $y$ and the per-frame estimate $y_t$.
Therefore, the estimate $y_t$ is only indirectly learned via back-propagation from the loss between the prediction $y$ and ground truth $\hat{y}$.

Our approach is located within a teacher-student framework, whereas a teacher $\mathcal{T}$ is first trained to estimate $y$.
After training, $\mathcal{T}$ then predicts soft-labels $\hat{y}_t$ on a known or unknown dataset, providing frame-level supervision to a student $\mathcal{S}$.
Note that in our work, the teacher is trained to predict $E$ (here $E = 527$, including ``Speech'' and  526 ``non-Speech'' events) different events, whereas the student $\mathcal{S}$ is trained as a binary classifier between speech and non-speech.

Therefore, the soft training labels $\hat{y}_{t}^{\mathcal{S}}$ for student $\mathcal{S}$ given the predictions $y_{t}^{\mathcal{T}}$ of teacher $\mathcal{T}$ are defined as:
\begin{align}\label{eq:label_pooling_student}
\begin{split}
    \hat{y}_{t}^{\mathcal{S}}(\text{Speech}) & = y_{t}^{\mathcal{T}}(\text{Speech})\\
    \hat{y}_{t}^{\mathcal{S}}(\text{non-Speech}) &= \max_{e \ne \text{Speech}} y_{t}^{\mathcal{T}}(e)
    \end{split}
\end{align}

Since the goal is to best discriminate between speech and non-speech events, we utilize the maximal value across all events not labeled as ``Speech'' (see \Cref{eq:label_pooling_student}) as the negative class (non-Speech) representation.
For the positive ``Speech'' class, we use the naive approach of directly transferring the teacher's predictions to the student.
Please note that $\hat{y}_t(\text{Speech}) + \hat{y}_t(\text{non-Speech}) \neq 1$, which enables our model to simultaneously predict speech, as well as possible foreground or background noises.
Also, during inference, we only consider the outputs of $y_t(\text{Speech})$ as being valid and neglect $y_t(\text{non-Speech})$.

\section{Experiments}
\label{sec:experiments}

In this section, we introduce the experimental setup, including utilized datasets for training and evaluation and insights about the used framework.
All neural networks were implemented in Pytorch~\cite{PaszkePytorch}.

\subsection{Datasets}

We first provide details on the training and evaluation datasets. 
All datasets' duration and data condition (clean, real) can be seen in \Cref{tab:datasets}.

\paragraph*{Training Data}
It should be noted that since we adopt a teacher-student approach, the training data utilized in this work is split into two categories: 
\begin{enumerate*}
  \item Source data, which is used to train a teacher model. The source data is labeled on the clip-level.
  \item Target data, which is unlabeled. The teacher is estimating frame-level soft labels on a target dataset. Then a student model is trained from scratch on this dataset and evaluated.
\end{enumerate*}
\paragraph{Source data}
In this work, we utilize the publicly available Audioset~\cite{Gemmeke2017} dataset for our backbone teacher training.
The commonly available Audioset is split into a ``balanced'' (further $\mathcal{A}_1$) and an ``unbalanced'' (further $\mathcal{A}_2$) subset. 
The ``balanced'' $\mathcal{A}_1$ dataset was collected by first taking examples for the rarest classes, then moving on to less-rare classes, ultimately leading to at least 59 examples for each event (but 5000+ for the most seen ``Music'' event).
The main difference between the $\mathcal{A}_1$ and $\mathcal{A}_2$ datasets is the amount of available data.
Due to difficulties obtaining the entire dataset, our $\mathcal{A}_1$ subset contains 21k, and $\mathcal{A}_2$ contains 1.85M at most 10-second long Youtube audio clips.
The data can be considered unconstrained since the dataset is taken from the globally utilized Youtube platform; thus, parameters such as recording devices, environment, data quality are unknown.
Audioset is annotated at clip-level, with 527 possible event classes, where it should be noted that label noise (e.g., incorrect labels) is present.
Within these 527 events, our focus lies in the ``Speech'' class event. 
The ``Speech'' event according to the Audioset ontology contains: ``Male speech'', ``Female speech'', ``Child speech'', ``Conversation'', ``Monologue'', ``Babbling'' and ``Synthesized speech''.
Unlike other datasets, Audioset is not restricted to one specific language, meaning that the teacher model can be considered language-agnostic.

\begin{figure}[htbp]
    \centering
    \includegraphics[width=0.95\linewidth]{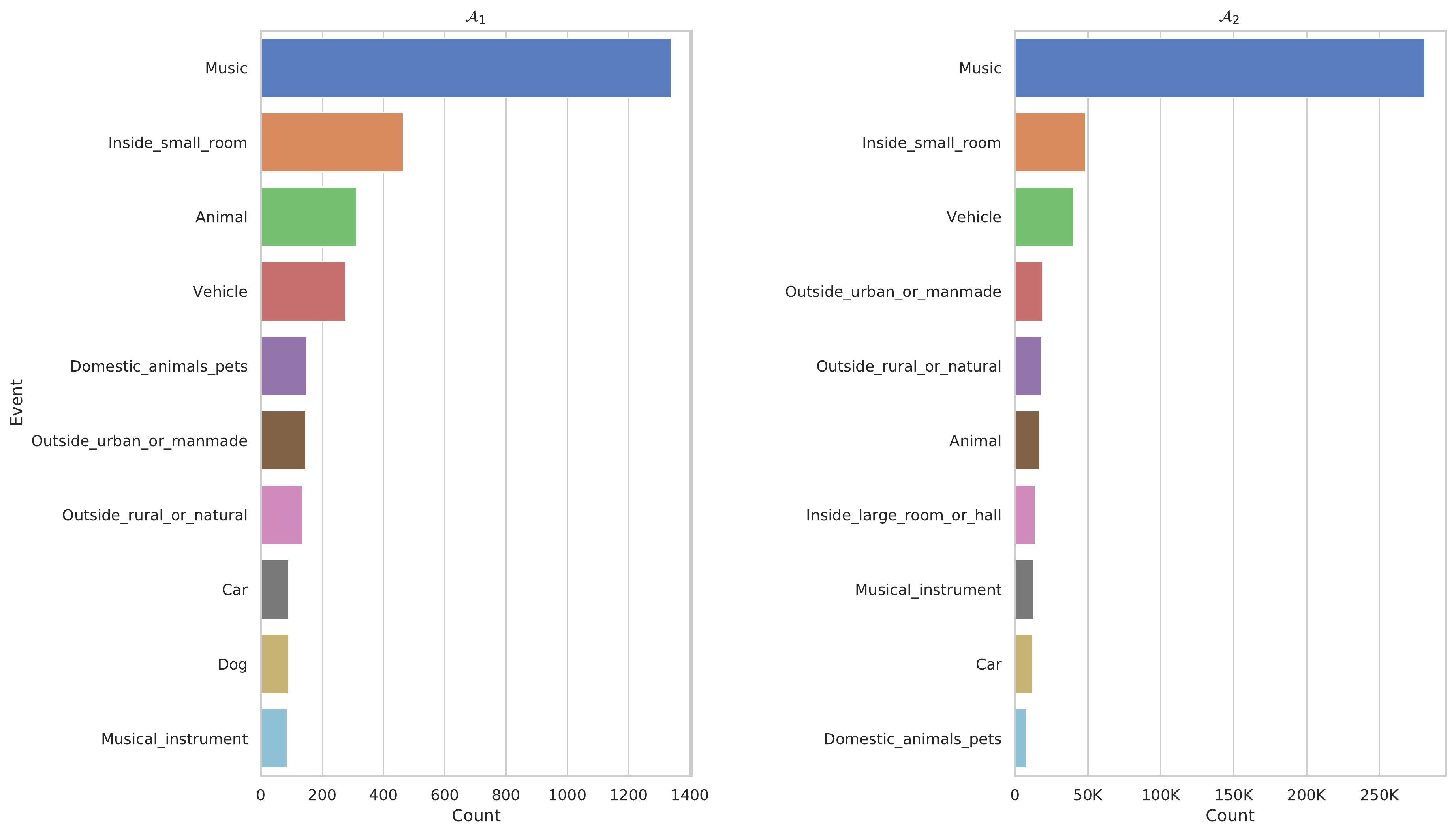}
    \caption{Top 10 most common ``non-Speech'' event co-occurring with ``Speech'' within $\mathcal{A}_{1}$ (left) and $\mathcal{A}_2$ (right) datasets.}
    \label{fig:occurance}
\end{figure}

The $\mathcal{A}_1$ subset contains 5452 clips ($\approx$ 15h), the $\mathcal{A}_2$ subset 905721 ($\approx$ 2500h) clips labeled as ``Speech''.
Note that $\mathcal{A}_1$ only contains samples, where ``Speech'' is seen with other events in tandem ($\mathbf{U}=\left(\mathbf{u}_1,\ldots,\mathbf{u}_L \right)$), whereas $\mathcal{A}_2$ contains single individual ``Speech'' only samples ($\mathbf{U}=\left(\mathbf{u}_0, \mathbf{u}_1,\ldots,\mathbf{u}_L \right)$).
The amount of events co-occurring with ``Speech'' in $\mathcal{A}_1$ is $L=405$, while for $\mathcal{A}_2$ it is $L=498$. 
Therefore, it is likely that training a teacher on $\mathcal{A}_2$ is potentially more noise-robust than on $\mathcal{A}_1$.
The most common events co-occurring with ``Speech'' for each respective dataset are provided in \Cref{fig:occurance}.

\paragraph{Target data}
The target data consists of the two datasets utilized for teacher training ($\mathcal{A}_{1/2}$), as well as three other datasets.
These three datasets are: Voxceleb1 ($\mathcal{V}_1$)~\cite{Nagraniy2017}, VoxCeleb2 ($\mathcal{V}_2$)~\cite{Chung2018,Nagrani2020}, as well as $\mathcal{V}_3$ which is a combination of the SRE datasets~\cite{sadjadi20172016} and Switchboard datasets~\cite{godfrey1992switchboard}.
$\mathcal{V}_{1/2}$ are collected from Youtube; thus, data can contain real-world noises but is likely only to contain spoken language as their primary sound source. $\mathcal{V}_1$ contains about 150,000 audio clips from more than 1200 speakers. The average length of audios is 8.2s, and the whole corpus contains approximately 352 hours of audio. The collection of $\mathcal{V}_2$ follows the same procedure as $\mathcal{V}_1$, but with many more speakers involved. 
About 1.13M audio clips from about 6000 speakers are contained in $\mathcal{V}_2$, with an average duration of 7.8s and a total duration of 2442 hours.

Unlike $\mathcal{A}_{1/2}$ and $\mathcal{V}_{1/2}$, which are collected from open-source Youtube videos, $\mathcal{V}_3$ was carefully planned and constructed by asking users to record phone calls. 
$\mathcal{V}_3$ consists of a Switchboard (SWBD) portion and an SRE portion, where the former contains SWBD phase 2,3 and cellular 1,2, and the latter contains NIST SRE04-10.
$\mathcal{V}_3$ is commonly used for the SRE challenges and contains long-duration recordings, with an average duration of 5 minutes. 
Overall, $\mathcal{V}_3$ contains more than 60000 recordings, leading to a total duration of 5213 hours.

\paragraph*{Evaluation Data}
Three different evaluation scenarios are proposed. 
First, we validate our model on the clean Aurora 4 test set (test A)~\cite{hirsch2000aurora}. 
Test A contains 330 utterances with a total duration of 40 minutes.
Second, we synthesize a noisy test set based on the clean Aurora 4 test set by randomly adding noise from a database of 100 noise files encompassing 20 noise types (e.g., Machine, Crowd, Traffic, Animal, Water, Cry, Laugh, Yawn) using an SNR ranging from 5db to 15db in steps of 1db (test B).
Lastly, we merge the development and evaluation tracks of the Challenge on Detection and Classification of Acoustic Scenes and Events 2018 (DCASE18)~\cite{Serizel2018}, itself a subset of Audioset, to create our real-world evaluation data (test C).
The DCASE18 data provides ten domestic environment event labels, of which we neglect all labels other than ``Speech'', but report the number of instances where non-speech labels were present.
The DCASE18 dataset contains manually re-annotated samples from Audioset, where similar event labels are generally merged (e.g., ``Cat'' + ``Meow'' from Audioset $\rightarrow$ ``Cat'' in DCASE18).
An important difference between DCASE18 and Audioset is that the manually annotated events in DCASE18 are comparatively noise-free, meaning that wrong (incorrect) or absent (incomplete) event labels are rarely seen, whereas Audioset contains label-noise.
Our DCASE18 evaluation set encompasses 596 utterances labeled as ``Speech'', 414 utterances (69\%) contain another non-speech label, 114 utterances (20\%) only contain speech, and 68 utterances (11\%) contain two or more non-speech labels.
This indicates that the test set (C) is the most challenging trial compared with (A) and (B).
We summarize the differences between the evaluation dataset as follows.
\begin{enumerate*}[label=\arabic*)]
\item Evaluation sets A and B are annotated using an automatic HMM alignment, whereas test C is manually annotated using human-labor.
\item Tests A and B contain exclusively English speech, whereas test C contains an unknown amount of languages. 
\item Test C contains sporadic speech (e.g., random shouts or greetings), whereas tests A, B only contains well-pronounced (e.g., news broadcast) sentences in English.
\end{enumerate*}

\begin{table}[htbp]
    \centering
    \begin{tabular}{ll|r|r|r|r}
    \toprule
        \multicolumn{2}{c}{Datatype} & Name   & Condition & Label & Duration  \\
        \midrule
        \multirow{5}{*}{\rotatebox[origin=c]{90}{Target}} & \multirow{2}{*}{Source} & Balanced ($\mathcal{A}_1$) & Real & Clip &  60 h \\
        & & Unbalanced ($\mathcal{A}_2$) & Real & Clip &  5000 h \\
        \cline{2-6}
        & & VoxCeleb1 ($\mathcal{V}_1$) & Real & - & 352 h\\
        & & VoxCeleb2 ($\mathcal{V}_2$) & Real & - & 2442 h\\
        & & SRE ($\mathcal{V}_3$)  & Clean & - & 5213 h \\
        \hline
        \multicolumn{2}{c|}{\multirow{3}{*}{Evaluation}}  & Aurora 4 (A) & Clean & Frame & 40 min \\
        & & Aurora 4 (B) & Syn & Frame & 8.7 h \\
        &  & DCASE18 (C) &  Real & Frame &  100 min \\
        \bottomrule
    \end{tabular}
    \caption{Training datasets for teachers (source) and students (target) as well as the three proposed evaluation protocols for clean, synthetic noise and real-world scenarios. Duration represents the overall duration of any signal in the corpus.}
    \label{tab:datasets}
\end{table}

\subsection{Setup}
Our VAD experiments used $64$-dimensional log-Mel power spectrograms (LMS) in this work regarding feature extraction.
Every single audio-clip is resampled to $22050$ Hz.
Each LMS sample was extracted by a $2048$ point Fourier transform every $20$ ms with a window size of $40$ ms using a Hann window.

\begin{equation}
    \label{eq:bce}
    \mathcal{L}(\hat{y}, y) = \hat{y} \log(y) + (1-\hat{y})\log(1-y) 
\end{equation}

The training criterion for all experiments between the ground truth $\hat{y}$ and prediction $y$ is binary cross-entropy (BCE, see\Cref{eq:bce}).
Regarding teacher training, the BCE is computed on clip-level, while student training computes BCE per frame.

Linear softmax~\cite{Wang2018,dinkel2019duration} (\Cref{eq:linear_softmax}) is utilized as temporal pooling layer ($\Gamma$) that merges frame-level probabilities $y_t(e) \in \left[ 0,1 \right]$ to a single vector representation $y(e) \in \left[ 0,1 \right]^E$.
\begin{equation}\label{eq:linear_softmax}
          y(e) = \frac{\sum_{t}^T y_t(e)^2}{\sum_{t}^T y_t(e)}
\end{equation}

Linear softmax is only utilized during teacher training and removed during student training.

\subsection{Evaluation metrics}
\label{ssec:metrics}

Our models are evaluated on two distinct levels: frame-level and segment-level.
All binary metrics used in this work require:
\begin{itemize}
    \item True positive (TP): Both reference and system prediction indicates speech to be present.
    \item False positive (FP): System prediction indicates speech to be present, but the reference indicates non-speech.
    \item False negative (FN): Reference indicates speech to be present, but the system prediction indicates non-speech.
\end{itemize}

\paragraph{Frame-level} For frame-level evaluation, we utilize macro averaged (instance-independent) precision (P), recall (R), and their corresponding F1 score.
Moreover, we also report the frame error rate (FER).
\begin{align}
\begin{split}
  \text{P} &= \frac{\text{TP}}{\text{TP}+\text{FP}}, \text{R} = \frac{\text{TP}}{\text{TP}+\text{FN}}\\
  \text{F1} &= 2\frac{\text{PR}}{\text{P}+\text{R}}\\
  \text{FER} &= \frac{\text{FP} + \text{FN}}{\text{TP}+\text{FP}+\text{FN}+\text{TN}}
  \label{eq:eval_metrics}
  \end{split}
\end{align}
The threshold-based metrics (P, R, F1, FER) can be seen in \Cref{eq:eval_metrics}.
Moreover, to compare different approaches with each other, independent of the post-processing or thresholds used, we also include Area Under the Curve (AUC)~\cite{ROC_AUC}. 
Note that the computation of AUC is directly done on the estimated speech probability sequence $y_t(\text{Speech}) \in [0,1]$.

\paragraph{Segment-level} For segment-level evaluation we utilize event-based F1-Score (Event-F1)~\cite{Mesaros2016,Bilen2019}.
Event-F1 calculates whether onset, offset, and the predicted label overlaps with the ground truth, therefore being a measure for temporal consistency.
We set a t-collar value according to WSSED research~\cite{Serizel2018} to 200 ms to allow an onset prediction tolerance and further permit a duration discrepancy between the reference and prediction of 20\%.

\subsection{Models}
\label{ssec:models}

Both teacher and student models utilize the same convolutional recurrent neural network (CRNN) back-end.
The architecture consists of a five-layer CNN (utilizing $3\times3$ convolutions), summarized into three blocks, with L4-Norm pooling after each block~\cite{dinkel2019duration,Dinkel2020a}, identical to the CDur framework from~\cite{Dinkel2021a}.
A bidirectional gated recurrent unit (BGRU) is attached after the last CNN output, enhancing our models' temporal consistency.
The framework and specific parameters can be seen in \Cref{fig:data_driven_vad}.
The model has 679k parameters, making it comparably light-weight, only requiring 2.7 MB on disk.

\paragraph*{Teacher model}
This work uses two teacher models, $\mathcal{T}_{1/2}$. 
$\mathcal{T}_{1}$ represents our baseline teacher approach, only utilizing the smaller $\mathcal{A}_1$ dataset with no augmentation, identical to the CRNN in~\cite{Dinkel2020a}.
Further, we propose $\mathcal{T}_2$, which is trained on the large $\mathcal{A}_2$ dataset, utilizing additional augmentation as seen in \Cref{ssec:augment} (SpecAug, Time shift).

To provide insight into our teacher models' potential performance implications, we evaluated on a subset ($\approx$ 36h) of the official Audioset evaluation data.
Please note that these results are computed on clip-level, meaning they have little importance for frame-level performance and can be viewed as a measure of our model's capability to estimate non-speech sound events.
The results in \Cref{tab:sourcemodel_performance} show that the additional training data for $\mathcal{T}_2$ training leads to better outcomes regarding mean average precision (mAP), AUC, and d-prime ($d^{'}$) compared to $\mathcal{T}_1$.  
However, the performance lacks behind large CNN models~\cite{Gemmeke2017,Kong2018a,Kong2019a,Xu2018}.
The main reason for this performance discrepancy is that our approach aims at modeling speech, which requires a high time-resolution, ultimately leading to poor clip-level performance.
The high time-resolution requirement also partially hinders our network's depth and width since our approach can not arbitrarily diminish the time-dimension.
Lastly, since VAD is a pre-processing step to other tasks, a fast run-time speed is generally preferred, meaning large models should be avoided.

\begin{table}[htbp]
    \centering
    \begin{tabular}{ll|rrrr}
        \toprule
        Source & Teacher & Aug & mAP & AUC & $d^{'}$ \\
        \midrule
        $\mathcal{A}_1$ & $\mathcal{T}_1$ & \xmark &  10.9 & 88.5 & 1.698  \\
        $\mathcal{A}_2$ & $\mathcal{T}_2$ & \cmark & 22.6 & 92.9 & 2.080 \\ 
        \bottomrule
    \end{tabular}
    \caption{Teacher models and their respective performance on the Audioset evaluation data. Only $\mathcal{T}_2$ utilizied augmentation during training.}
    \label{tab:sourcemodel_performance}
\end{table}

\paragraph*{Student model}

The student model is structurally identical to the teacher model (see \Cref{fig:data_driven_vad}).
Unlike the teacher model, the student is trained on the teacher's frame-level predictions and does not require the temporal pooling function $\Gamma$.
BCE is utilized as the frame-level loss function (\Cref{eq:bce}).

\subsection{Training}

Teacher training mainly differs from student training with its data sampling strategy.
We utilize a balanced data sampling strategy aiming at oversampling minority sound events such that each batch at best contains a single sample per sound event.
Note since this is a multi-label classification problem, perfect label balance is impossible since a minority event sample might also contain a majority event.
All student models are trained on 90\% of the available training data and cross-validated using the leftover 10\%.
Training of $\mathcal{T}_2$ slightly differs from this train/cross-validation paradigm, in which case we utilize the $\mathcal{A}_1$ dataset for cross-validation.
VAD training is done using Adam optimization with a starting learning rate of $0.001$, where the learning rate is reduced by a factor of $10$ if no improvement on the held-out cross-validation set has been seen for at least $5$ cross-validation steps.
The batch-size for all experiments was set to $64$.
Cross-validation is done after each epoch or every 5000 batches.
For all utilized datasets, training is run for 15 epochs.
The best model obtaining the lowest loss on the held-out cross-validation dataset is kept for inference/evaluation.
During training, zero-padding is applied to each audio-clip towards the longest clip's length within a batch.
Since our student models observe frame-level labels, we mask our loss such that each padded element does not influence the final back-propagation step.
Code and pretrained models are available online \footnote{Available at github.com/richermans/datadriven-GPVAD}.

\subsection{Augmentation}
\label{ssec:augment}

The training utilizes the following data augmentation schemes.

\paragraph{SpecAug}
Recently, a cheap yet effective augmentation method has been introduced named SpecAugment (SpecAug)~\cite{park2019specaugment}.
SpecAug randomly sets time-frequency regions to zero within an input log-Mel spectrogram.
Time modification is applied by masking $\gamma_{t}$ times $\eta_{t}$ consecutive time frames, where $\eta_{t}$ is chosen to be uniformly distributed between $[t_0, t_0 + \eta_{t0}]$ and $t_{0}$ is uniformly distributed within $\left[0, T-\eta_{t} \right)$.
Frequency modification is applied by masking $\gamma_f$ times $\eta_{f}$ consecutive frequency bins $\left[f_0, f_0 + \eta_{f}\right)$, where $\eta_{f}$ is randomly chosen from a uniform distribution in the range of $\left[0, \eta_{f0}\right]$ and $f_0$ is uniformly chosen from the range $\left[0, D-\eta_{f}\right)$. 
When using SpecAug, we set $\gamma_t = 2, \eta_{t0} = 60, \gamma_{f} = 2, \eta_{f0} = 8$.
Note that SpecAug is utilized during teacher ($\mathcal{T}_2$) as well as any student training.

\paragraph{Time Shifting}

Time shifting is utilized only during teacher training since it does not affect student training (frame-level labels). 
Since only clip-level labels are present, we encourage the model to learn time-coherent predictions.
For each audio-clip, we draw $\eta_{sh}$ from a normal distribution $\mathcal{N}(0, 10)$, meaning that we randomly either shift the audio clip forward or backward by $\eta_{sh}$ frames.

\subsection{Post-processing}
\label{ssec:post-processing}
During evaluation, post-processing is required to obtain hard labels from class-wise probability sequences ($y_t(e)$).
We hereby use double threshold~\cite{dinkel2019duration,Kong2018b} post-processing, which uses two thresholds $\phi_{\text{low}}=0.1, \phi_{\text{hi}}=0.5$.
Please note that double thresholding aims to enhance the temporal consistency, therefore being beneficial in terms of Event-F1.

\section{Results And Analysis}
\label{sec:vad_results}

In this section, we provide our experimental results and insight into the possible limits of our method.
Please note that we consider FER, Event-F1, and AUC as our primary metrics, whereas P, R, and F1 are considered secondary metrics.

\subsection{Baseline}
\label{ssec:base}

Here we first introduce our baseline approaches. 
First, we compare our clip-level trained teachers to a frame-level trained VAD-C (CRNN) model from~\cite{Dinkel2020a}.
The VAD-C model back-end is identical to our CRNN framework, where only the training data (Aurora 4) and supervision (frame-level) differ, and artificial noise is added during training.
Therefore, our VAD-C baseline is an example of a traditional supervised VAD approach with clean training data.


\begin{table}
\centering
\begin{tabular}{ll||rrrrrrrr}
\toprule
     Test & Model & P & R & F1  & AUC & FER & Event-F1 \\
     \midrule
     \multirow{3}{*}{A} &  VAD-C & \textbf{97.97} & \textbf{95.32} & \textbf{96.55}  & \textbf{99.78} & \textbf{2.57} & \textbf{78.90} \\
     &  $\mathcal{T}_1$ & 95.69 & 95.47 & {95.58}  &  {99.07} &  {4.01} &  {73.70} \\
     &  $\mathcal{T}_2$ & 96.13 & 94.97 & {95.52}  &  {97.75} &  {3.38} &  {70.10} \\
     \hline
     \multirow{3}{*}{B}  & VAD-C &  \textbf{91.37} & {82.82} & {85.96} & \textbf{97.07} & \textbf{9.71} & \textbf{47.50} \\
     &  $\mathcal{T}_1$ & 80.34  & 87.58 & {81.99}  &  {94.63} &  {15.74} &  {35.40} \\
     &  $\mathcal{T}_2$ & 85.28 &  \textbf{90.25} & \textbf{87.17}  &  {95.22} &  {10.58} &  {42.50} \\
     \hline
     \multirow{3}{*}{C} & VAD-C &   78.17 & 79.08  & 77.93  & 87.87 & 21.92 &  {34.40} \\
     & $\mathcal{T}_1$ & 85.36 & 82.70 & {83.50} &  {91.80} & {15.47} & {44.80} \\
     &  $\mathcal{T}_2$ & \textbf{88.15} & \textbf{86.21} & \textbf{86.89}  &  \textbf{94.58} &  \textbf{12.74} &  \textbf{46.30} \\
     \bottomrule
\end{tabular}
\caption{A baseline comparison between traditional supervised (CRNN) VAD-C approach trained on Aurora 4 in a frame-supervised manner to our proposed teacher models.}
\label{tab:baseline_results}
\end{table}

The results can be seen in \Cref{tab:baseline_results}.
Unsurprisingly, VAD-C outperforms our proposed clip-level training teachers on the clean (A) and synthetic (B) test sets.
However, the difference in performance between our teachers and VAD-C is acceptable since our approach has no strong frame-level supervision.

Leveraging large-data ($\mathcal{A}_2$) for $\mathcal{T}_2$ shows promising performance against $\mathcal{T}_1$ when noise is present.
In real-world scenarios (test C), both $\mathcal{T}_1$ and $\mathcal{T}_2$ significantly outperform the standard VAD model on all shown metrics.
Specifically, we observe a significant drop in FER (21.92 $\rightarrow$ 12.74) and an increase in AUC (87.87 $\rightarrow$ 94.58).
Further, note that performance is less affected by noise (compare results B and C), indicating noise robustness for both our teacher models.

\subsection{Difference in label types}
\label{ssec:label_type_train}

Naturally, since the teacher model outputs probabilities (\textit{soft labels}), an interesting topic of investigation is if \textit{hard labels} (zero-one) are helpful during training. 
We believe that both soft- and hard-label approaches mutually benefit each other.
We assume that hard labels are possibly beneficial to detect onset- and offset boundaries. In contrast, soft labels can more effectively provide duration estimates since speech to non-speech transitions are smooth.
We conduct two experiments, using each respective teacher model $\mathcal{T}_{1/2}$.
Three types of labels are utilized: 
\begin{itemize}
    \item Soft labels, i.e., probabilities, $\hat{y}_t^{\mathcal{S}} \in [0, 1]$ (soft).
    \item Hard labels obtained from thresholding all soft labels with $\phi=0.5$, $\hat{y}^{\mathcal{S}}_t \in \{0, 1\}$ (hard).
    \item Randomly (hard) thresholding at most $25\%$ of the speech samples within an audio-clip using $\phi=0.5$ (dynamic). Note that during our model selection phase, we investigated the thresholds $10\%, 25\%, 50\%$, and came to the conclusion that $25\%$ works best.
\end{itemize}

\begin{table}
\begin{tabular}{p{0.23cm}l||rrrrrr}
\toprule
Test & Label &  P &  R &    F1 &   AUC &   FER &  Event-F1 \\
\midrule
\multirow{4}{*}{A} & clip ($\mathcal{T}_1$) & 95.69 & 95.47 & {95.58}  &  {99.07} &  {4.01} &  {73.70}\\
        \cline{2-8}
        & hard &      94.18 &   \textbf{96.31} & 95.17 & \textbf{99.31} &  3.79 &     \textbf{76.34} \\
        & soft &      \textbf{96.07} &   95.03 & \textbf{95.53} & 98.93 &  \textbf{3.38} &     65.22 \\
        & dynamic &      94.06 &   94.98 & 94.50 & 98.56 &  4.25 &     61.19 \\
        \hline
\multirow{4}{*}{B} & clip ($\mathcal{T}_1$) & 80.34  & 87.58 & {81.99}  &  {94.63} &  {15.74} &  {35.40}\\
        \cline{2-8}
        & hard &      79.84 &   87.81 & 81.07 & 96.10 & 16.91 &     36.20 \\
        & soft &      \textbf{86.24} &   \textbf{90.80} & \textbf{88.04} & \textbf{96.70} &  \textbf{9.78} &     \textbf{48.79} \\
        & dynamic &      81.95 &   88.75 & 83.85 & 95.67 & 13.88 &     34.34 \\
        \hline
\multirow{4}{*}{C} & clip ($\mathcal{T}_1$) & 85.36 & 82.70 & {83.50} &  {91.80} & {15.47} & {44.80} \\
        \cline{2-8}
        & hard &      \textbf{85.39} &   80.97 & 82.00 & \textbf{93.12} & 16.55 &     48.75 \\
        & soft &      84.96 &   \textbf{83.86} & \textbf{84.28} & 92.70 & \textbf{15.01} &     \textbf{51.29} \\
        & dynamic &      83.70 &   81.86 & 82.47 & 90.63 & 16.58 &     44.84 \\
\bottomrule
\end{tabular}
\caption{Results using teacher $\mathcal{T}_1$ and student model trained on $\mathcal{A}_1$ using different label types. We compare the teacher $\mathcal{T}_1$ baseline using clip-level training to the student models. Best results are highlighted in bold.}
\label{tab:soft_vs_hard_frame_level_t1}
\end{table}

Our initial results using the baseline teacher $\mathcal{T}_1$, and student model trained on $\mathcal{A}_1$ can be seen in \Cref{tab:soft_vs_hard_frame_level_t1}.
Here, we compare the students trained on frame-level using the three proposed label types against the clip-level teacher $\mathcal{T}_1$.
First and foremost, it can be seen that our proposed teacher-student approach improves performance in all test scenarios (A, B, C) against the teacher $\mathcal{T}_1$ (clip).
For example, the AUC from $\mathcal{T}_1$ increases on test A (99.07 $\rightarrow$ 99.31), B (94.63 $\rightarrow$ 96.70) and C (91.80 $\rightarrow$ 93.12) when using teacher-student training.
Second, our results indicate that hard-label training is preferred in clean data scenarios to obtain consistent temporal predictions.
Here, the hard-label approach on the test set A improves the Event-F1 score from 73.70 to 76.34.
This observation seems to be in line with our baseline VAD-C method, which is also trained on hard-labels, as it is common for traditional VAD approaches.

\begin{table}
\begin{tabular}{p{0.23cm}l|rrrrrr}
\toprule
Test & Label          &  P &  R &    F1 &   AUC &   FER &  Event-F1 \\
\midrule
\multirow{4}{*}{A} & clip ($\mathcal{T}_2$) & 96.13 & \textbf{94.97} & \textbf{95.52}  &  {97.75} &  {3.38} &  \textbf{70.10}\\
\cline{2-8}
& hard &      96.70 &   93.57 & 95.00 & \textbf{98.65} &  3.71 &     66.39 \\
        & soft &      \textbf{97.23} &   {94.06} & {95.51} & 98.26 &  \textbf{3.33} &     {69.19} \\
        & dynamic &     97.05 &   93.60 & 95.16 & 98.50 &  3.57 &     66.25 \\
        \hline
\multirow{4}{*}{B} & clip ($\mathcal{T}_2$) & 85.28 &  90.25 & 87.17  &  {95.22} &  {10.58} &  {42.50} \\
\cline{2-8}
 & hard &      88.42 &   90.09 & 89.20 & 96.86 &  8.45 &     52.84 \\
        & soft &      \textbf{90.60} &   90.81 & 90.70 & 97.23 &  7.14 &     \textbf{57.78} \\
        & dynamic &     90.44 &   \textbf{91.32} & \textbf{90.87} & \textbf{97.37} &  \textbf{7.08} &     55.29\\
        \hline
\multirow{4}{*}{C} & clip ($\mathcal{T}_2$) & {88.15} & {86.21} & {86.89}  &  {94.58} &  {12.74} &  {46.30}\\
\cline{2-8}
        & hard &      87.55 &   85.63 & 86.30 & 94.99 & 12.98 &     \textbf{55.35} \\
        & soft &      86.78 &   85.46 & 85.96 & \textbf{95.14} & 13.38 &     54.72 \\
        & dynamic &     \textbf{88.19} &   \textbf{86.79} & \textbf{87.33} & 95.02 & \textbf{12.08} &     54.91 \\
\bottomrule
\end{tabular}
\caption{Results using teacher $\mathcal{T}_2$ and student model trained on $\mathcal{A}_1$ using different label types. We compare the teacher $\mathcal{T}_2$ baseline using clip-level training to the student models. Best result per test in bold.}
\label{tab:soft_vs_har_frame_level_t2}
\end{table}

Further, we also provide our results using teacher $\mathcal{T}_2$ in \Cref{tab:soft_vs_har_frame_level_t2}.
Student models are also trained on $\mathcal{A}_1$.
The performance increase from using the more potent teacher $\mathcal{T}_2$ is evident in the noisy test cases (B, C).
All frame-level and segment-level metrics improve significantly compared to the teacher model, e.g., on test B, FER $10.58 \rightarrow 7.08$ and Event-F1 $42.50 \rightarrow 57.78$. 
Moreover, while both teachers perform worse than our baseline VAD-C approach on test B, the students of teacher $\mathcal{T}_2$ now outperform VAD-C in both B and C noisy test-conditions regarding AUC, FER, and Event-F1.
Different from the previous observations in \Cref{tab:soft_vs_hard_frame_level_t1}, it seems that our dynamic labeling method is consistently superior to soft and hard-label approaches for test B and C in regards to FER and F1.
Lastly, our performance gap across the synthetic trial B and real noise trial C scenarios is significantly less than the VAD-C baseline.
Notably, on all tested conditions, our AUC is higher than $95$, indicating our approach's noise robustness.
Due to the results in \Cref{tab:soft_vs_har_frame_level_t2}, all further experiments utilize by default teacher $\mathcal{T}_2$ and use the dynamic label-scheme (for further information, see \Cref{sec:soft_vs_dynamic_appendix}).

\subsection{Teacher-student VAD using unlabeled out-of-domain data}
\label{ssec:mixed_data_vad}

One of our approach's significant advantages is that it can potentially scale to other, out-of-domain datasets. 
Since the teachers are trained on real-world data, they can provide frame-level supervision on any dataset without being constrained to any specific data type (clean, real) or other conditions such as language.
So far, our work utilized Audioset ($\mathcal{A}_1$) to achieve substantial improvements in noisy environments but slightly lag behind the clean test A.
This performance gap could stem from the large amount of non-speech events within Audioset. 
We hypothesize that adding data that mainly contains speech (e.g., $\mathcal{V}_1$) might be beneficial.
Thus, this experiment mainly focuses on comparing $\mathcal{V}_1$ and $\mathcal{A}_1$ as target datasets.

\begin{table}
\begin{tabular}{p{0.75cm}l|rrrrrr}
\toprule
Target & Test    &  P &  R &    F1 &   AUC &   FER &  Event-F1 \\
\midrule
\multirow{3}{*}{$\mathcal{A}_1$} & A & 97.05 &   93.60 & 95.16 & {98.50} &  3.57 &     66.25 \\
& B & 90.44 &   91.32 & 90.87 & 97.37 &  7.08 &     55.29 \\
& C & 88.19 &   86.79 & 87.33 & 95.02 & 12.08 &     54.91\\
\hline
\multirow{3}{*}{$\mathcal{V}_1$} & A  &  96.64 &   94.36 & 95.43 & 98.07 &  3.42 &    70.62  \\
& B  &  91.77 &   90.20 & 90.94 & 96.56 &  6.81 &     55.60 \\
& C  & 87.28 &   86.94 & 87.10 & 94.41 & 12.45 &     53.47 \\
\hline
\multirow{3}{*}{$\mathcal{V}_1+\mathcal{A}_1$} & A  &  96.72 &   {94.74} & {95.67} & 98.45 & {3.24} &     69.43 \\
& B  &     90.49 &   91.36 & 90.91 & 97.24 &  7.04 &     54.16 \\
& C  &      85.98 &   85.41 & 85.66 & 94.41 & 13.79 &     54.57 \\
\bottomrule
\end{tabular}
\caption{Student training using the largely speech-only dataset $\mathcal{V}_1$ in conjunction with the noisy $\mathcal{A}_1$. Teacher $\mathcal{T}_2$ is utilized to predict dynamic labels on each respective dataset.}
\label{tab:mixed_data_results}
\end{table}

Our results in \Cref{tab:mixed_data_results} show that our approach can achieve competitive performance even when trained on other datasets (here $\mathcal{V}_1$).
Performance on the clean (A) test set improves against the $\mathcal{T}_2$ baseline.
We believe that the performance improvement in our (B) evaluation dataset stems from the possible language match (English) when training ($\mathcal{V}_{1}$).
Interestingly, by training on clean datasets ($\mathcal{V}_{1}$), performance in noisy test scenarios does not drop compared to real-world datasets ($\mathcal{A}_{1}$, see \Cref{tab:soft_vs_har_frame_level_t2}).
We assume that this is due to the teacher's noise-robust soft labels, indicating a knowledge transfer from teacher to student.
Adding real-world data to clean data (e.g., $\mathcal{V}_1 + \mathcal{A}_1$) seems to perform worse on the test set C than training on both datasets individually.

\subsection{Scaling with large data}

As it has been already seen in \Cref{ssec:base} and \Cref{ssec:label_type_train}, results of $\mathcal{T}_2$ substantially outperform results from $\mathcal{T}_1$, likely being due to the inherently much larger data size for teacher training.
We further investigate the implications of large target data utilization ($\mathcal{A}_2,\mathcal{V}_2$, $\mathcal{V}_3$) for student training.

\begin{table}
    \begin{tabular}{p{0.75cm}l|rrrrrr}
\toprule
Target & Test &  P &  R &    F1 &   AUC &   FER &  Event-F1 \\
\midrule
\multirow{3}{*}{$\mathcal{A}_2$}&  A & 96.96 &   94.28 & 95.52 & 98.13 &  3.34 &     69.10 \\
& B & 89.65 &   90.96 & 90.27 & 96.82 &  7.58 &     54.82\\
& C & 87.71 &   87.66 & 87.68 & 94.89 & 11.92 &     54.09\\
\hline
\multirow{3}{*}{$\mathcal{V}_2$} & A &      96.80 &   94.34 & 95.48 & 98.26 &  3.37 &     70.95 \\
& B &      91.71 &   89.79 & 90.69 & 96.65 &  6.97 &     55.43 \\
& C &     86.32 &   86.92 & 86.56 & 94.20 & 13.15 &     53.16 \\
\hline
\multirow{3}{*}{$\mathcal{V}_2+\mathcal{A}_2$} & A &      96.53 &   94.77 & 95.61 & 98.31 &  3.30 &     73.00 \\
& B &      89.62 &   91.47 & 90.48 & 97.19 &  7.47 &     53.96 \\
& C &      87.96 &   87.27 & 87.56 & 94.98 & 11.94 &     55.26 \\
\hline
\multirow{3}{*}{$\mathcal{V}_3$} & A &      96.84 &   95.10 & 95.93 & 98.66 &  3.06 &     74.80 \\
& B  &      90.44 &   92.87 & 91.54 & 97.63 &  6.68 &     54.45 \\
& C  &      \textbf{89.20} &   \textbf{88.36} & \textbf{88.72} & \textbf{95.20} & \textbf{10.82} &     \textbf{57.85} \\
\bottomrule
\end{tabular}
\caption{Large data training on labels generated from teacher $\mathcal{T}_2$ for each respective dataset using the dynamic labeling scheme. Our best (most noise robust across noisy evaluation scenarios) model is highlighted in bold.}
\label{tab:scaling_with_data}
\end{table}

Our results in \Cref{tab:scaling_with_data} demonstrate that our teacher-student approach can scale with data.
Our method can also be extended to any dataset, while some differences are notable between different target data.

The best performing model we have observed is trained on the $\mathcal{V}_3$ dataset since it achieved the lowest FER (3.06), as well as the highest Event-F1 (74.80) of any proposed teacher-student approaches, on the clean test A scenario.
More importantly, on both noisy tests (B, C), this model outperforms other models trained on $\mathcal{V}_{1/2}$ data, as well as $\mathcal{A}_2$.
Most importantly, this model achieves the highest performance on our difficult C trial.
Compared to our strong teacher $\mathcal{T}_2$ baseline, we observe an absolute decrease in FER by 3.9\%, an increase in AUC by 2.41, and an increase in Event-F1 by 11.95\% on test B.
While a relative improvement on test C is less than on test B (likely due to harder difficulty), the model still manages to decrease FER by 1.92\%, increase AUC by 0.68, and Event-F1 by 11.55\%. All metrics are reported in absolute.

\begin{figure}
    \centering
    \subfloat[Aurora 4 Noisy (B)]{%
       \includegraphics[width=0.9\linewidth]{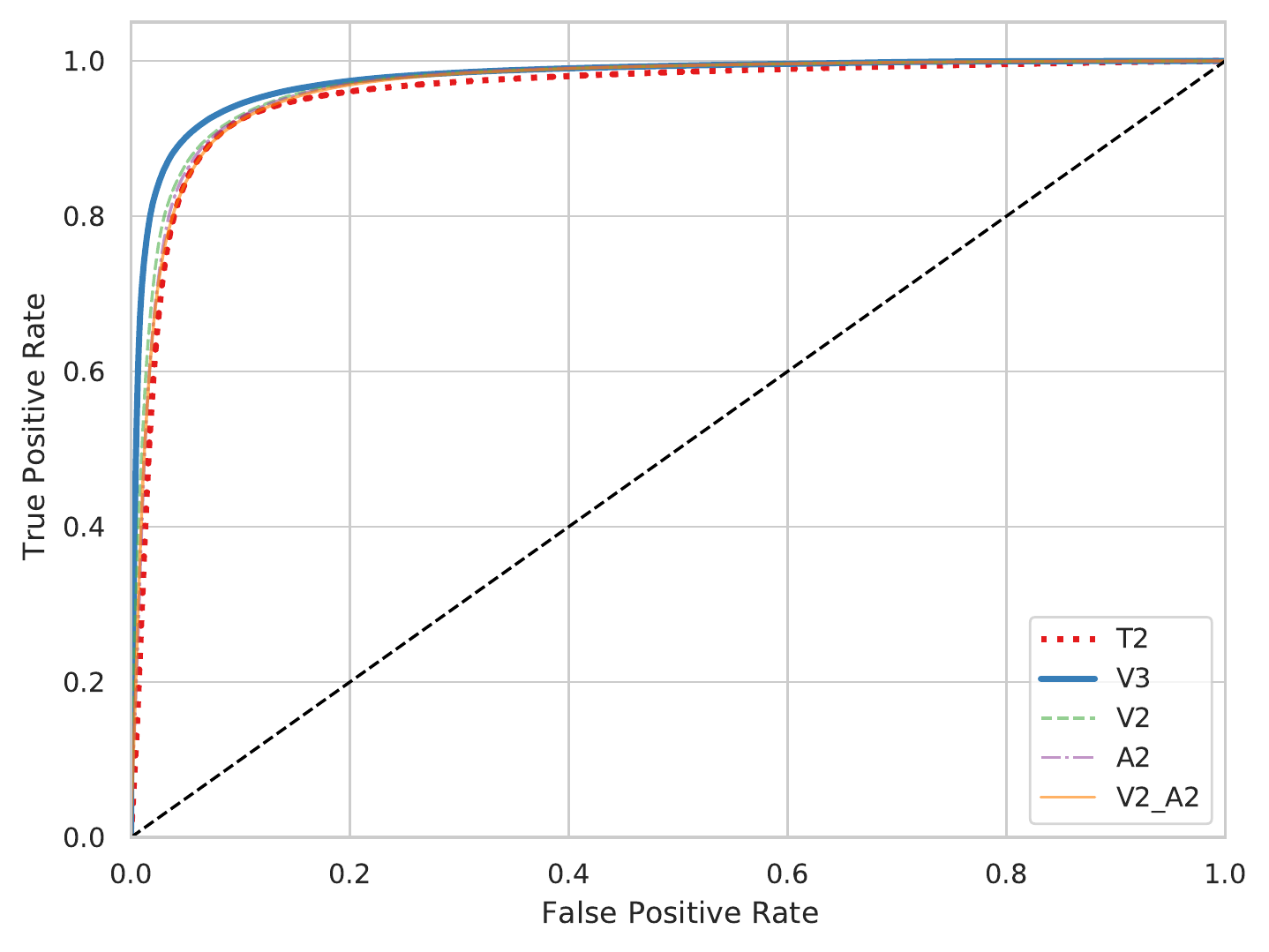}}
    \\
      \subfloat[DCASE18 (C)]{%
            \includegraphics[width=0.9\linewidth]{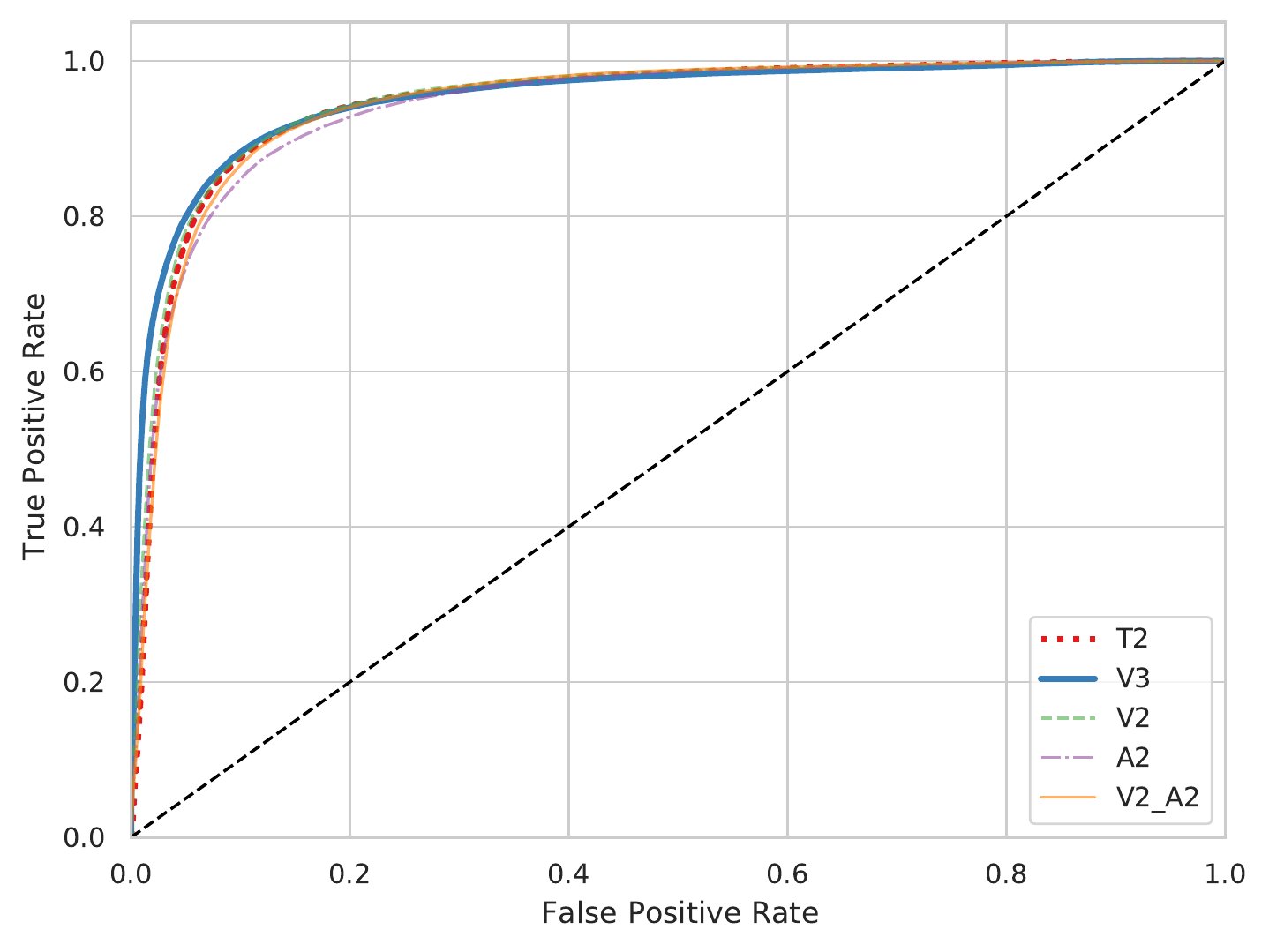}}
    \caption{Receiver operating characteristic (ROC) curves for the Aurora 4 Noisy (B) and DCASE18 evaluation sets. The teacher $\mathcal{T}_2$ is compared to its students $\mathcal{V}_2,\mathcal{A}_2,\mathcal{V}_2 + \mathcal{A}_2,\mathcal{V}_3$. Best viewed in color.}
    \label{fig:roc_plots_teacher_vs_student}
\end{figure}

Lastly, we also provide the receiver operating characteristic (ROC) curves for the results in \Cref{tab:scaling_with_data} in \Cref{fig:roc_plots_teacher_vs_student}.
Specifically, the ROC curves for the teacher $\mathcal{T}_2$ and its students are displayed.
We limit our visualization to tests B and C since the performance on those tests differs the most.

\subsection{Data size vs. target data characteristics}

Another essential question worth investigating is whether the previous results of the $\mathcal{V}_3$ model stems exclusively from the increased data size compared to $\mathcal{V}_2$ or if the reason is the characteristics of the $\mathcal{V}_3$ dataset.
For this reason, we subsampled the previously used $\mathcal{V}_3$ dataset to be of equal size to the $\mathcal{V}_2$ dataset, i.e., around 2400 hours.
Then, we trained a new teacher ($\mathcal{V}_3$ (2.4k)) on this subset, and the results can be observed in \Cref{tab:v3_vs_subsampled_v3}.

From the results, it can be noted that:
\begin{enumerate*}
    \item The new $\mathcal{V}_3$ (2.4 k) model performs well against the other approaches on the clean test A and obtains the highest Event-F1, AUC, and Recall results.
    \item Both $\mathcal{V}_3$ and $\mathcal{V}_3$ (2.4k) outperform $\mathcal{V}_2$ on test C.
\end{enumerate*}

\begin{table}
\centering
    \begin{tabular}{p{0.85cm}l|rrrrrr}
\toprule
Target & Test &  P &  R &    F1 &   AUC &   FER &  Event-F1 \\
\midrule
\multirow{3}{*}{$\mathcal{V}_2$} & A &      96.80 &   94.34 & 95.48 & 98.26 &  3.37 &     70.95 \\
& B &      91.71 &   89.79 & 90.69 & 96.65 &  6.97 &     55.43 \\
& C &     86.32 &   86.92 & 86.56 & 94.20 & 13.15 &     53.16 \\
\hline
\multirow{3}{*}{$\mathcal{V}_3$} & A &      96.84 &   95.10 & 95.93 & 98.66 &  3.06 &     74.80 \\
& B  &      90.44 &   92.87 & 91.54 & 97.63 &  6.68 &     54.45 \\
& C  &      \textbf{89.20} &   \textbf{88.36} & \textbf{88.72} & \textbf{95.20} & \textbf{10.82} &     \textbf{57.85} \\
\hline
\multirow{3}{*}{$\mathcal{V}_3$ (2.4k)} & A &      96.46 &   95.46 & 95.46 & 98.94 &  3.07 &     77.10 \\
& B  &      88.79 &   93.06 & 90.55 & 97.54 &  7.66 &     46.90 \\
& C  &      {89.11} & {87.18} & {87.87} & {94.88} & 11.49 &     {55.40} \\
\bottomrule
\end{tabular}
\caption{Comparison of the 2400 hour long $\mathcal{V}_2$ data against the subsampled $\mathcal{V}_3$ (2.4k). Best results for the hard C trial are highlighted in bold.}
\label{tab:v3_vs_subsampled_v3}
\end{table}

We conclude from these results that the dataset's size for student training is less important than its characteristics.
Comparing the results using medium-sized datasets in \Cref{tab:mixed_data_results} to the large scale ones in \Cref{tab:v3_vs_subsampled_v3} leads to the conclusion that, while larger datasets possibly contain more content-rich data, the performance benefits are marginal.
Instead, our approach seems to be well suited for cross-domain adaptation.

\subsection{Performance under different SNRs}
\label{ssec:snr_performance}

Here we further analyze the performance of our approach under synthetic noise scenarios.
A new noise-controlled test set is generated by mixing the clean audio from the test set A by noise from Musan~\cite{musan2015} in a range between 20 to -5 dB SNR (in steps of 5 dB).
Musan contains three categories of noise: speech, music, and background noise.
For each sample in the test set A, we independently add speech, music, and background noise, resulting in a test set three times the size of A for each SNR value.

\begin{figure*}
    \centering
    \includegraphics[width=\linewidth]{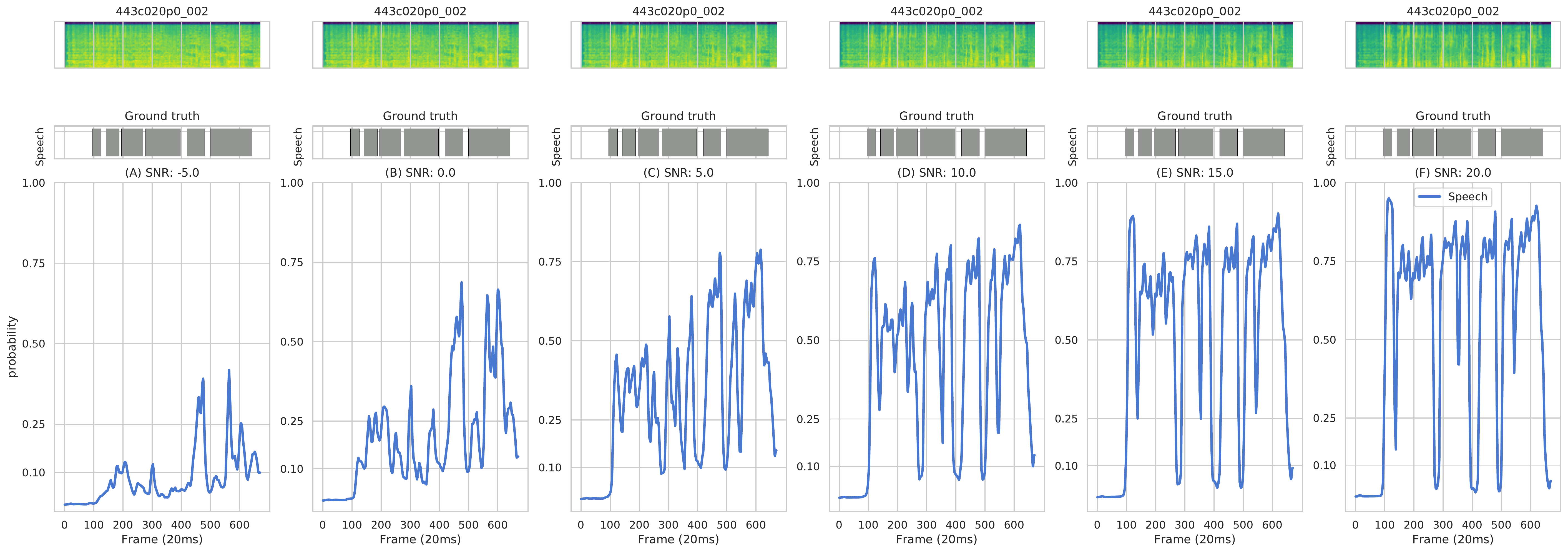}
    \caption{Our best model ($\mathcal{V}_3$) predicting speech under different SNRs ranging from -5 (A) to 20 (F) db in steps of 5db. Each plot title is a respective sample name from the Aurora4 dataset. Each graph represents a log-Mel spectrogram (top), ground truth (center) and probability output (bottom). Noise is exclusively music. }
    \label{fig:best_model_snr_plot_music}
\end{figure*}

\begin{figure*}
    \centering
    \includegraphics[width=\linewidth]{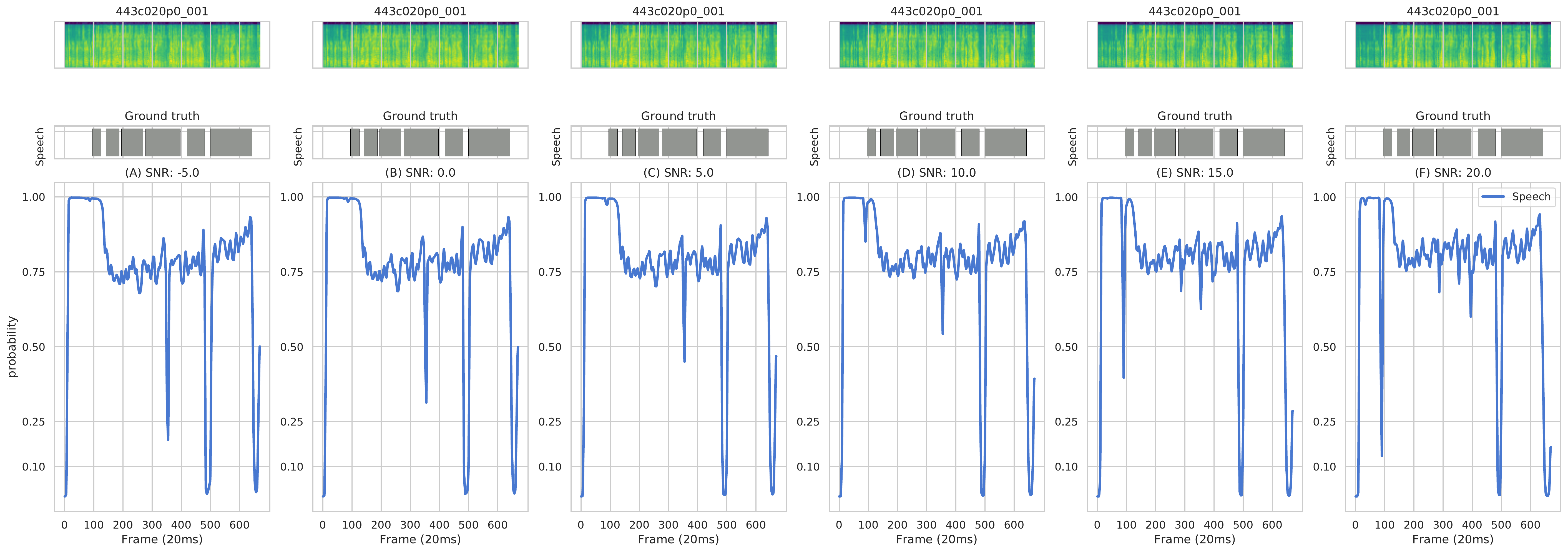}
    \caption{Our best model ($\mathcal{V}_3$) predicting speech under different SNRs ranging from -5 (A) to 20 (F) db in steps of 5db. Each plot title is a respective sample name from the Aurora4 dataset. Each graph represents a log-Mel spectrogram (top), ground truth (center) and probability output (bottom). Noise is exclusively speech.}
    \label{fig:best_model_snr_plot_speech}
\end{figure*}

\begin{figure*}
    \centering
    \includegraphics[width=\linewidth]{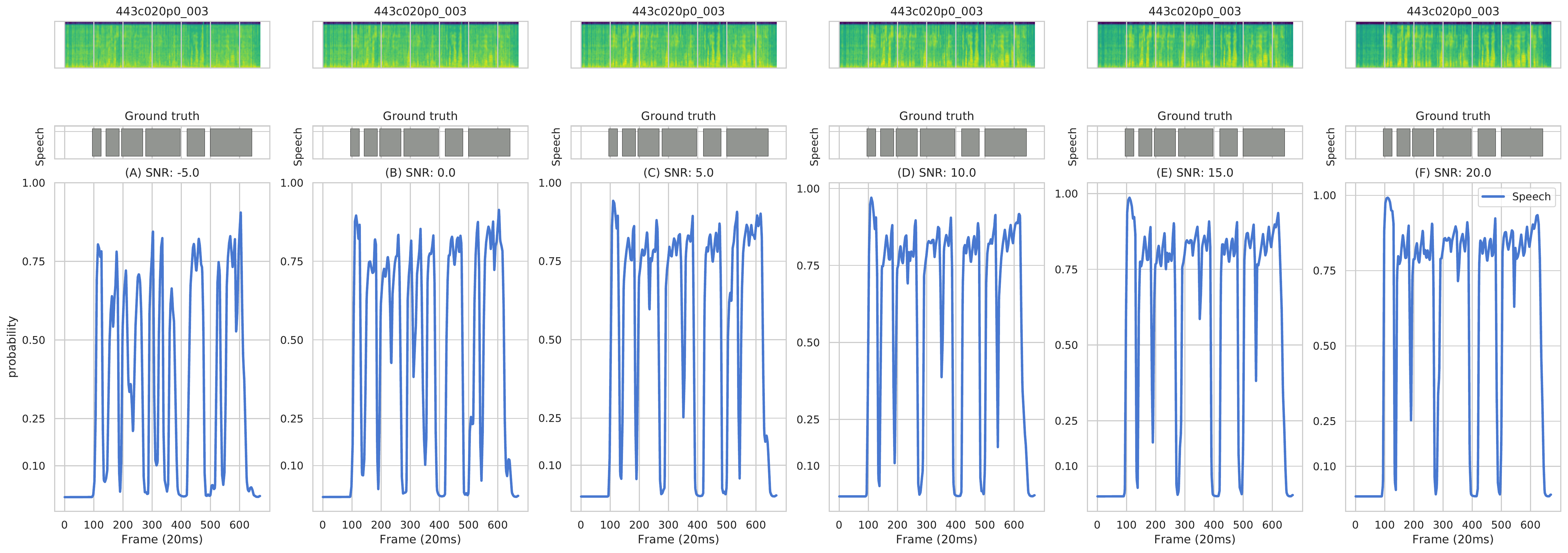}
    \caption{Our best model ($\mathcal{V}_3$) predicting speech under different SNRs ranging from -5 (A) to 20 (F) db in steps of 5db. Each plot title is a respective sample name from the Aurora4 dataset. Each graph represents a log-Mel spectrogram (top), ground truth (center) and probability output (bottom). Noise is exclusively background noises.}
    \label{fig:best_model_snr_plot_background_noise}
\end{figure*}

As the results indicate in \Cref{tab:snr_analysis}, our proposed approach is robust to noise, capable of providing adequate performance (FER 12.30, Event-F1 36.75) in noisy SNR = 0db scenarios.
However, it seems that our model's performance severely degrades even in light noise conditions (SNR = 20db).
We hypothesize that this increase stems from the additive speech noise, which would inevitably lead to our VAD's false activations.
We provide insight into our approach's limits via visualization of our models' probabilities for a comparatively hard sample, where speech occurred six times in a span of 12 s.
Here the individual samples utilizing music (\Cref{fig:best_model_snr_plot_music}), speech (\Cref{fig:best_model_snr_plot_speech}), and background noise (\Cref{fig:best_model_snr_plot_background_noise}) can be observed.

\begin{table}
\centering
\begin{tabular}{l|rrrrrr}
\toprule
             SNR          &  P &  R &    F1 &   AUC &   FER &  Event-F1 \\
\midrule
-5 &      76.50 &   78.83 & 77.48 & 81.63 & 18.04 &     28.10 \\
0  &      84.54 &   82.57 & 83.47 & 84.17 & 12.30 &     36.75 \\
5 &   84.36 & 86.55 & 85.90 & 85.90 &  9.64 &     45.21 \\
10 &      91.91 &   85.14 & 87.82 & 87.27 &  8.61 &     51.58 \\
15 &      92.93 &   85.67 & 88.52 & 88.63 &  8.09 &     55.94 \\
20 &      93.49 &   86.16 & 89.04 & 90.37 &  7.72 &     56.65 \\
Clean &  96.84 &   95.10 & 95.93 & 98.66 &  3.06 &     74.80 \\
\bottomrule
\end{tabular}
\caption{Our best model ($\mathcal{V}_3$) evaluated on the Aurora 4 corpus with additive noise (music, speech, background) from musan.}
\label{tab:snr_analysis}
\end{table}

First, as it can be seen in \Cref{fig:best_model_snr_plot_music}, our approach excels at noisy background scenarios such as music.
For SNR values of \textgreater 5db, it can be observed that the model is capable of effectively predicting speech boundaries and the presence of speech.
Notably, with high SNR values, the AUC of our approach can reach up to 95\%, which in turn decreases with a decrease in SNR.
However, for the hard SNR = -5 case, our approach, even though capable of sensing speech, does only output probabilities below 50\%, meaning that a change in post-processing would be useful (e.g., lower threshold).

Second, when faced with additional speech in \Cref{fig:best_model_snr_plot_speech}, our approach's potential drawbacks are observed. 
Our model now consistently outputs with high confidence in the presence of speech.
The prediction patterns produced appear to be very similar, regardless of SNR.
Even at SNR = 20, speech is predicted with high confidence throughout the entire utterance, indicating our model's high sensitivity towards speech.
This high sensitivity attests that our method is fully capable of detecting the presence of any speech. 
However, it currently cannot distinguish between, e.g., multiple speakers or different sound sources.
However, since our work can be easily extended with speaker-dependent VAD approaches, future work can focus on utilizing methods such as~\cite{Ding2020}.

Third, when confronted with common background noises in \Cref{fig:best_model_snr_plot_background_noise}, our approach shows little to no influence even under heavy noise (SNR = -5) scenarios, indicated by high probability values.
For all samples, it can also be seen that our model excels at estimating short, spontaneous bursts of speech, with accurate onset and offset prediction capabilities.

\subsection{Comparison with other approaches}

To prove our approach's effectiveness and the difficulty of VAD in real-world scenarios, we compare our results with previous successful frameworks.
Note that we use the default configuration of each proposed method. 
Thus input feature-types (i.e., MFCC) and hyper-parameters (i.e., frameshift) for all other approaches differ from ours.
Further, all other approaches were not retrained on our Aurora4 dataset and taken as-is from their respective public repository.
First, we compare our method to the naive energy-thresholding method used in the Kaldi~\cite{povey2011kaldi} toolkit.
Second, we utilize rVAD~\cite{Tan2020} (the rVAD-fast implementation), an unsupervised VAD approach, which has been seen to perform well in the presence of substantial noise.
Third, we also compare to traditional supervised VAD approaches using deep neural networks (DNN) from~\cite{Segbroeck2013}.
Lastly, we compare against a more modern attention-based approach (ACAM)~\cite{Kim2018}.
Note that our goal in this comparison is to show that previous approaches trained on their respective dataset cannot generalize to unseen noise types.
However, back-end models such as ACAM could be used in the future in conjunction with our proposed GPVAD approach to enhance performance further.
Additionally, since all other competitors' outputs are hard labels $y_t \in {0,1}$, we refrain from calculating the AUC score, denoted as ``--''.

\begin{table}
\centering
\begin{tabular}{p{0.23cm}l||rrrrrrrr}
\toprule
Test & Model & P & R &F1  & AUC & FER & Event-F1 \\
     \midrule
\multirow{6}{*}{A} &  VAD-C & \textbf{97.97} & \textbf{95.32} & \textbf{96.55}  & \textbf{99.78} & \textbf{2.57} & \textbf{78.90} \\
                   & Kaldi & 90.14 & 94.42 & 91.93 & - & 6.48 & 2.30 \\
                   & rVAD & 95.75 & 95.27 & 95.50 & - & 3.40 & 76.10 \\
                   & DNN & 87.24 & 93.75 & 89.52 & - & 8.74 & 27.10 \\
                   & ACAM & 96.38 & 89.96 & 92.61 & - & 5.26 & 55.20 \\
                   & Ours &  96.84 &   95.10 & 95.93 & 98.66 &  3.06 &     74.80 \\
                   \hline
\multirow{6}{*}{B}  & VAD-C & \textbf{91.37} & {82.82} & {85.96} & 97.07 & 9.71 & 47.50 \\
                    & Kaldi & 76.40 & 54.29 & 51.56 & - & 23.82 & 1.60 \\
                    & rVAD & 89.31 & 77.77 & {81.41} & - & 12.23 & 36.60 \\
                    & DNN & 79.69 & 87.36 & 81.21 & - & 16.50 & 11.40 \\
                    & ACAM & 89.25 & 84.50 & 86.50 & - & 9.71 & 35.80  \\
                    & Ours & 90.44 &   \textbf{92.87} & \textbf{91.54} & \textbf{97.63} &  \textbf{6.68} &     \textbf{54.45} \\
     \hline
\multirow{6}{*}{C} & VAD-C & 78.17 & 79.08 & 77.93  & 87.87 & 21.92 &  {34.40} \\
                   & Kaldi & 66.86 & 52.79 & 35.88 & - & 55.30 & 9.20 \\
                   & rVAD & 74.73 & 73.87 & 70.88 & -  & 29.07 & 39.80 \\
                   & DNN & 72.38 & 72.89 & 71.35 & - & 28.59 & 24.00\\
                   & ACAM & 73.19 & 70.96 & 66.99 & - & 32.75  & 12.30   \\
                   & Ours &  \textbf{89.20} &   \textbf{88.36} & \textbf{88.72} & \textbf{95.20} & \textbf{10.82} &     \textbf{57.85} \\
     \bottomrule
\end{tabular}
\caption{Comparison between traditional energy (Kaldi), unsupervised (rVAD), supervised (DNN, ACAM) and our baseline (VAD-C) approaches to our student model trained on $\mathcal{V}_3$ (Ours). Best achieved result per test is highlighted in bold.}
\label{tab:compare_other_approaches}
\end{table}

The results in \Cref{tab:compare_other_approaches} show that our chosen VAD-C baseline is indeed more potent than other approaches on clean data.
Standard Kaldi energy-thresholding offers a comparatively well-rounded performance on test A in terms of FER (6.48) and F1 (91.93) while profoundly lacking temporal consistency (Event-F1 2.30).
However, when noise increases (B, C), the naive Kaldi approach degenerates to random guessing levels (FER 55.30, F1 35.88).
Further, we observe that rVAD performs well in clean and synthetic noise scenarios, as seen in its original work~\cite{Tan2020}. 
However, when faced with real-world, unconstrained evaluation, its performance decreases significantly on test C.
Our proposed method shows signs of noise robustness between trials B and C, obtaining a lower FER and higher F1 in test C than rVAD does in test B.
Traditional supervised VAD models, only using a shallow 2-layer DNN structure from~\cite{Segbroeck2013} are unable to perform well even against the unsupervised rVAD approach.
Also, more modern attention-based approaches from~\cite{Kim2018} are seen to perform better than the traditional shallow DNN model.
However, both supervised approaches perform consistently worse than our VAD-C baseline, suggesting that our model architecture (CRNN) is indeed suited for supervised VAD.
In this comparison, it can be seen that our method is the best performing in noisy scenarios.
More importantly, its performance across multiple test scenarios is also the most stable (e.g., FER increases from 3.06 in test A to 10.82 in test C, and AUC drops from 98.66 to 95.20).

\begin{figure}
    \centering
    \subfloat{%
       \includegraphics[width=0.97\linewidth]{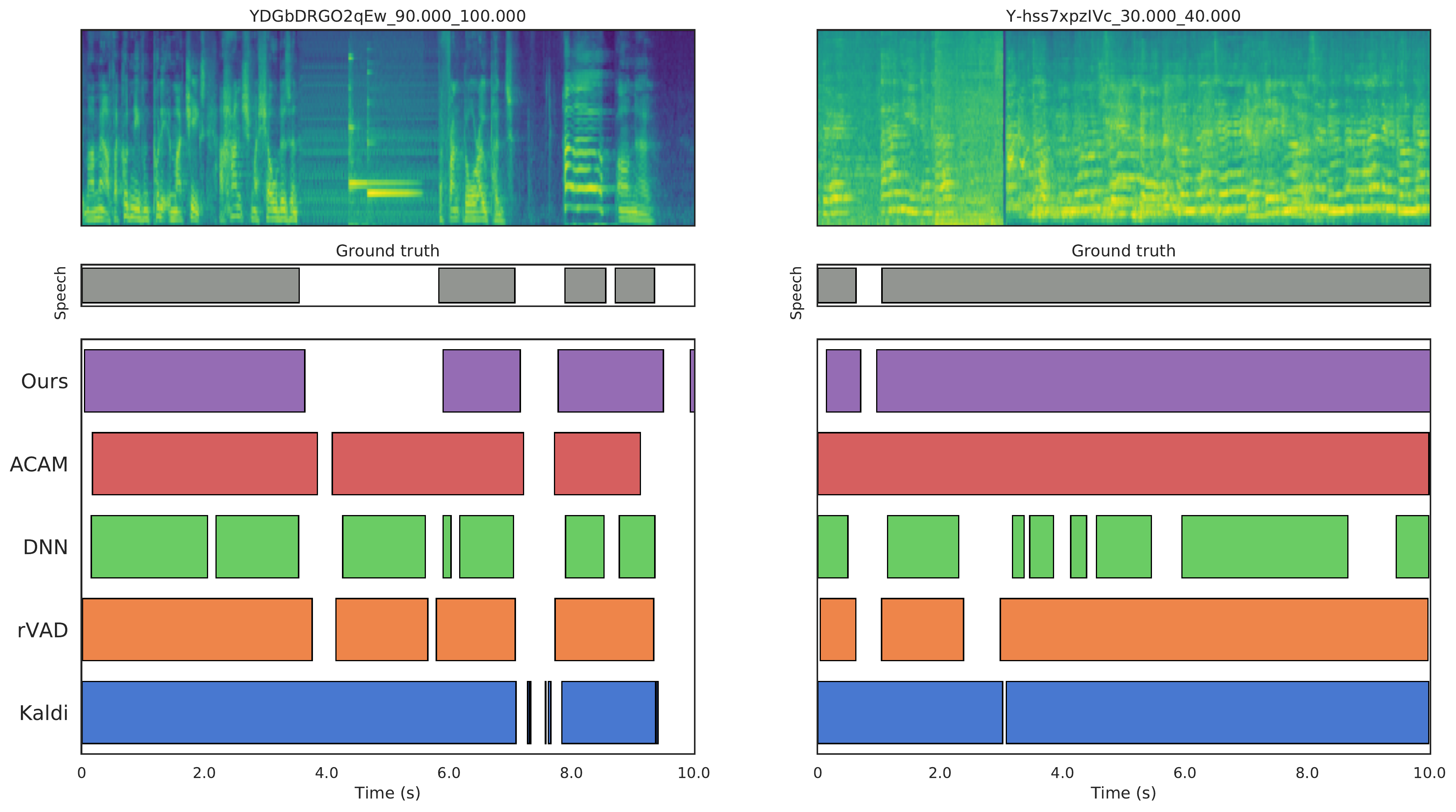}}
    \\
      \subfloat{%
            \includegraphics[width=0.97\linewidth]{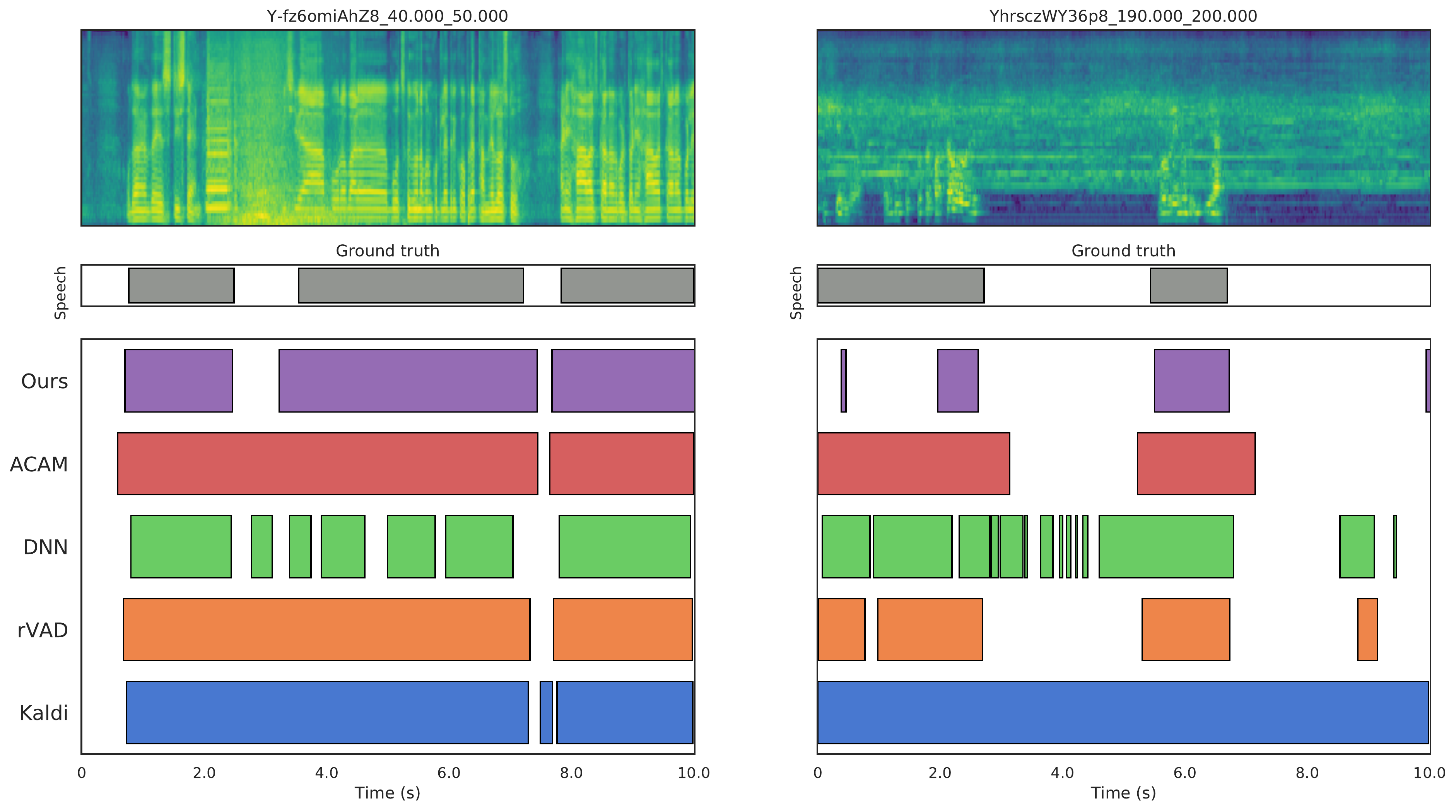}}
        \\
      \subfloat{%
        \includegraphics[width=0.97\linewidth]{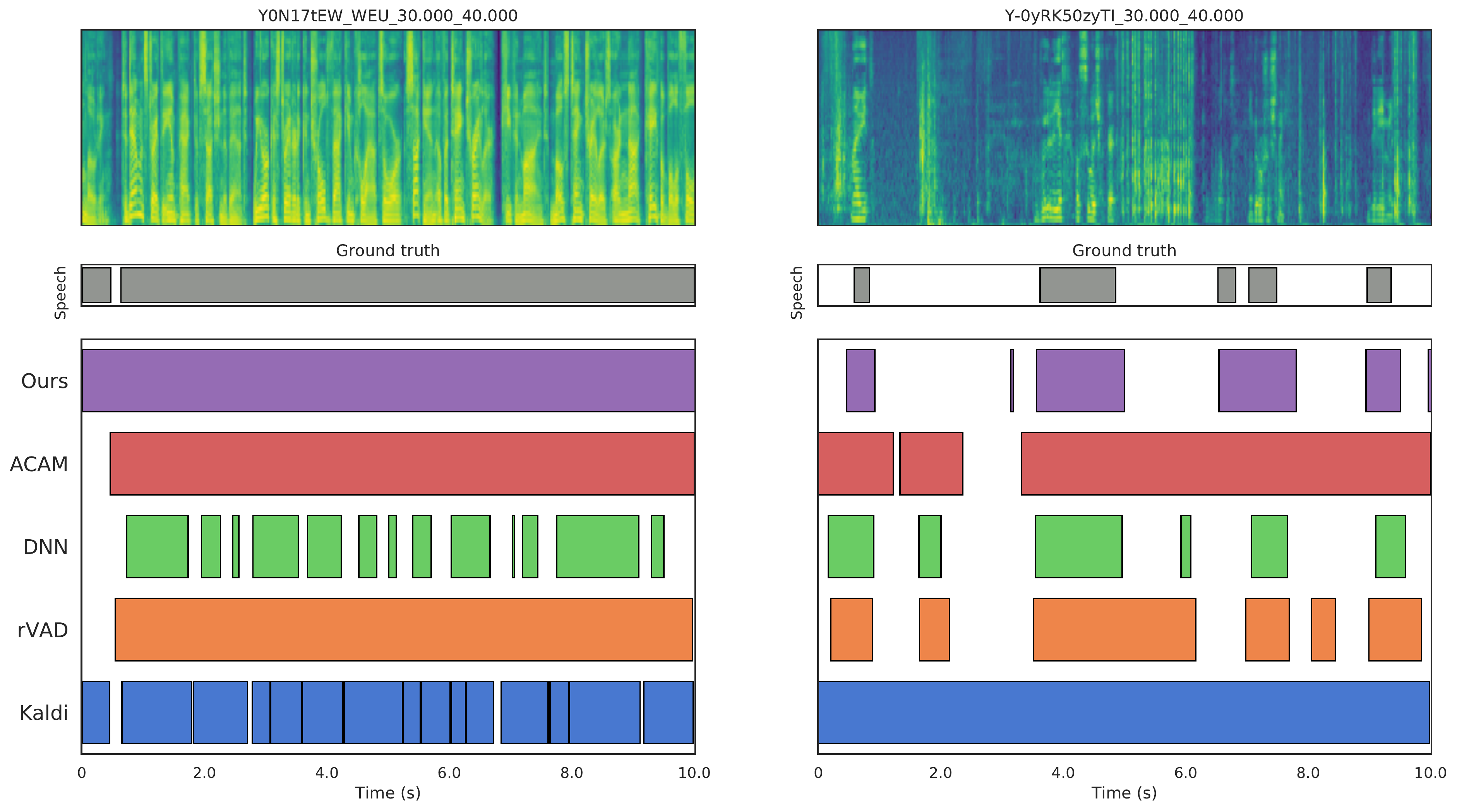}}\\
    \subfloat{%
        \includegraphics[width=0.97\linewidth]{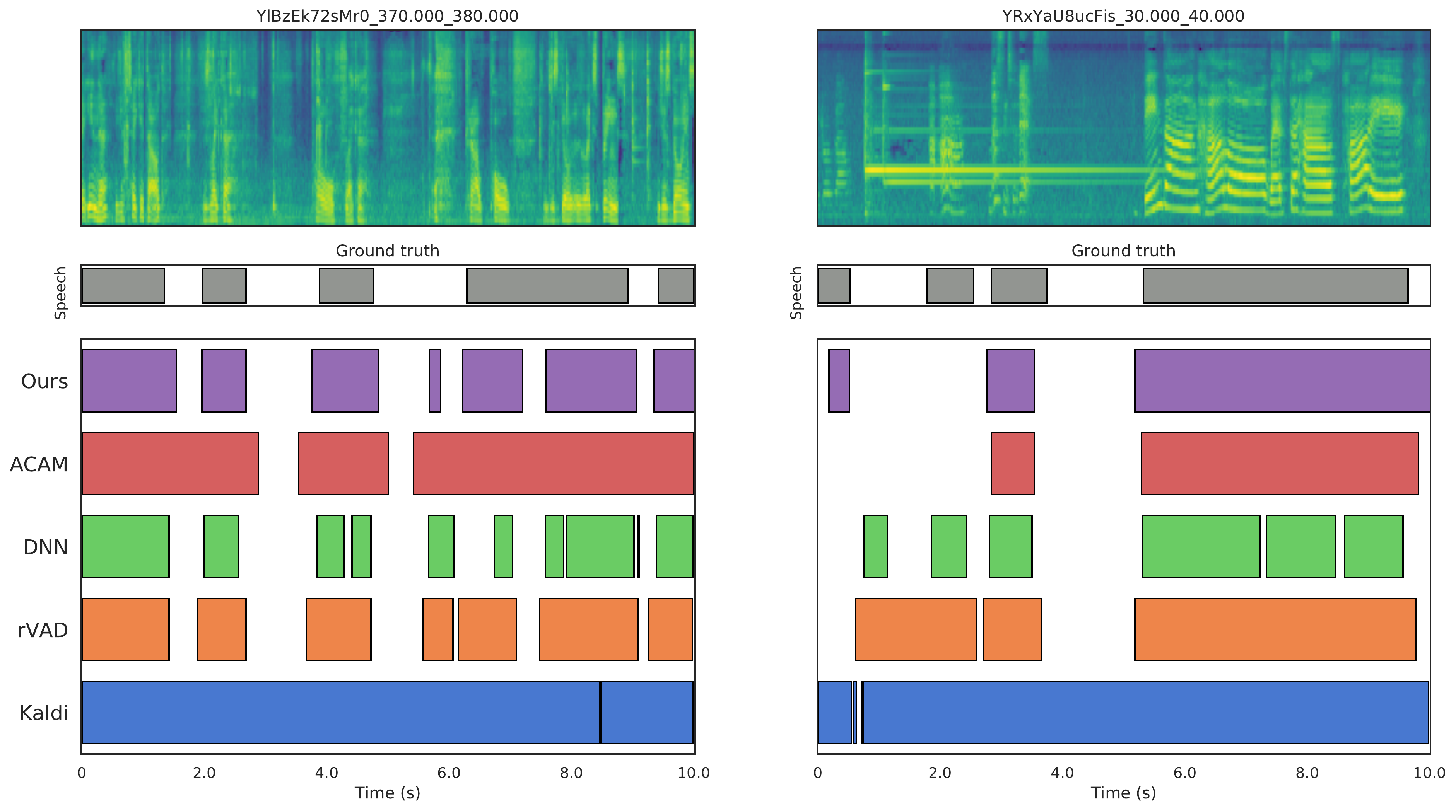}}    
    \caption{Eight sample predictions of our best student model ($\mathcal{V}_3$) using default post-processing against previous methods on test C. For each graph: (Top) LMS. (Center) Ground truth label. (Bottom) speech presence predictions in color for each respective model. Each plot title is a respective sample from the DCASE18 dataset (formatted as Y[Youtubeid\_start\_end]). Viewers are encouraged to visit each respective Youtube link for a better experience.}
    \label{fig:sample_comparison_1}
\end{figure}

We also visualize some sample predictions on trial C in \Cref{fig:sample_comparison_1} between all used models.
Note that since the test C labels are human-annotated, incorrect labeling can occur (e.g., short pauses are not considered).
Compared to other approaches, the visualizations demonstrate our model's superiority in terms of FER since it is rarely seen to mispredict speech activity.
Onsets (start of speech) and offsets (end of speech) are well estimated, even though our approach never had access to strong supervision (as the other comparable supervised models), thus needed to learn duration estimation by itself.
However, our model is also seen to miss out on predicting speech activity, which leads us to investigate its sensitivity.

\subsection{Sensitivity}

Here we study the post-processing impact on our model's sensitivity.
As per default, we used double thresholding (see \Cref{ssec:post-processing}), which can be seen as a conservative post-processing method.
For this experiment, we remove double thresholding as post-processing method and replace it with traditional thresholding $y_t^{\mathcal{S}}(\text{Speech}) > \phi$, where we investigate $\phi \in \{ 0.01,0.02,0.05, 0.1, 0.2, 0.5, 0.7 \}$. 

Here, False alarm rate ($P_{fa}$) is the percentage of ``non-Speech'' frames being mis-classified as ``Speech'' and ($P_{miss}$) is the percentage of ``Speech'' frames being misclassified as ``non-Speech''.
We compare our findings with other methods from \Cref{tab:compare_other_approaches}, utilizing their respective default configuration.
The results can be seen in \Cref{tab:sensitivity}, reflecting our previous findings regarding our method's noise robustness.
If a low threshold $0.01$ is used to reduce $P_{miss}$ to as low as $3.36$\%, the percentage of false accepts $P_{fa}$ still outperforms all other comparable approaches.
Also, note that for all investigated thresholds, our highest reported FER (20.58\%) remains lower compared to other approaches (see \Cref{tab:sensitivity}).

\begin{table}[htbp]
    \centering
    \begin{tabular}{ll|rrr}
    \toprule
        Method & Threshold ($\phi$) & $P_{fa}$\% & $P_{miss}$\% & FER\% \\
        \midrule
        \multirow{8}{*}{Ours} & default & 7.02 & 16.26  & 10.82  \\
        & 0.01 & 32.64 & 3.36 & 20.58 \\
        & 0.02 & 23.07 & 5.13 & 15.68 \\
        & 0.05 & 13.97 & 8.71 & 11.80 \\
        & 0.1 & 9.28 & 12.53  & 10.62 \\
        & 0.2 & 5.92 & 17.96  & 10.88 \\
        & 0.5 & 2.33 & 31.04 & 14.15 \\
        & 0.7 & 1.15 & 43.64 & 18.65 \\
        \hline
        DNN~\cite{Segbroeck2013} & - & 35.62 & 18.59 & 28.59\\
        ACAM~\cite{Kim2018} & - & 50.29 &  7.78 & 32.75\\ 
        Kaldi~\cite{povey2011kaldi} & - & 93.52 & 0.91 & 55.30 \\
        rVAD~\cite{Tan2010} & 0.7 & 31.40 & 19.88 & 26.65 \\
        rVAD~\cite{Tan2020} & default (0.4) & 42.98 & 9.29 & 29.07 \\
        rVAD~\cite{Tan2020} & 0.1 & 61.89 & 3.92 & 37.97\\
        \bottomrule
    \end{tabular}
    \caption{Sensitivity in regards of $P_{fa}$ and $P_{miss}$ as well as FER on the test C. Here default represents double thresholding with $\phi_{low}=0.1,\phi_{hi}=0.5$.}
    \label{tab:sensitivity}
\end{table}

\begin{table*}[htbp]
    \centering
    \begin{tabular}{lll|rrrrrr}
\toprule
Target & Task & Label  &  P & R &    F1 &   AUC &   FER &  Event-F1 \\
\midrule
\multirow{6}{*}{$\mathcal{V}_1$} & \multirow{2}{*}{A} & soft &      97.17 &   93.89 & 95.38 & 97.97 &  3.42 &     69.12 \\
        & & dyn &      96.64 &   94.36 & 95.43 & 98.07 &  3.42 &     70.62 \\
& \multirow{2}{*}{B} & soft &      91.88 &   89.04 & 90.33 & 96.40 &  7.17 &     56.00 \\
        & & dyn &      91.77 &   90.20 & 90.94 & 96.56 &  6.81 &     55.60 \\
& \multirow{2}{*}{C} & soft &      84.90 &   85.64 & 85.15 & 94.11 & 14.56 &     40.57 \\
        & & dyn &      87.28 &   86.94 & 87.10 & 94.41 & 12.45 &     53.47 \\
        \hline
\multirow{6}{*}{$\mathcal{V}_2$} & \multirow{2}{*}{A} & soft &      97.20 &   94.24 & 95.60 & 98.50 &  3.27 &     71.33 \\
       & & dyn &      96.80 &   94.34 & 95.48 & 98.26 &  3.37 &     70.95 \\
& \multirow{2}{*}{B} & soft &      91.62 &   89.73 & 90.62 & 96.64 &  7.03 &     55.84 \\
        & & dyn &      91.71 &   89.79 & 90.69 & 96.65 &  6.97 &     55.43 \\
& \multirow{2}{*}{C} & soft &      85.30 &   85.44 & 85.36 & 94.45 & 14.21 &     42.37 \\
        & & dyn &      86.32 &   86.92 & 86.56 & 94.20 & 13.15 &     53.16 \\
\hline
\multirow{6}{*}{$\mathcal{V}_1+\mathcal{A}_1$} & \multirow{2}{*}{A} & soft &      97.19 &   93.64 & 95.24 & 98.42 &  3.51 &     61.62 \\
        & & dyn &      96.72 &   94.74 & 95.67 & 98.45 &  3.24 &     69.43 \\
& \multirow{2}{*}{B} & soft &      92.39 &   90.43 & 91.35 & 97.40 &  6.48 &     57.02 \\
        & & dyn &      90.49 &   91.36 & 90.91 & 97.24 &  7.04 &     54.16 \\
& \multirow{2}{*}{C} & soft &      84.85 &   84.92 & 84.88 & 94.59 & 14.66 &     39.77 \\
    &    & dyn &      85.98 &   85.41 & 85.66 & 94.41 & 13.79 &     54.57 \\

\bottomrule
\end{tabular}
    \caption{Performance difference between soft and dynamic (dyn) labels on target data.}
    \label{tab:dynamic_vs_hard_appendix}
\end{table*}


\section{Conclusion}
\label{sec:conclusion}

This work proposes and investigates a novel data-driven teacher-student approach for voice activity detection to be trained with vast amounts of data.
A teacher model is firstly trained using clip-wise labels on Audioset.
Then the teacher is used to predict probabilities (soft-labels) for a student model.
In our initial results, we show that teacher-student training on both source datasets ($\mathcal{A}_{1/2}$) significantly benefits VAD performance in noisy test conditions.
Further, we investigate the influence of soft, hard, and dynamic labels on performance.
Our proposed dynamic approach is seen to outperform both soft and hard label training in noisy scenarios.
Out-of-domain large data student training is also investigated, utilizing the Voxceleb 1/2 datasets as well as NIST SRE.
Our best student model significantly outperforms our supervised VAD-C baseline as well as our teachers ($\mathcal{T}_{1/2})$ on all noisy evaluation scenarios regarding FER, F1, AUC, and Event-F1 metrics.
Notably, Event-F1 scores of over 50\% are reported across all test cases, meaning that our model excels at segmentation by providing accurate speech on- and offsets.
When comparing our method to traditional supervised and unsupervised approaches, noise robustness is observed in the difficult C trial.
The noise robustness is validated by our model's performance in the MUSAN corrupted A trial for low SNRs.
Moreover, our model is sensitive to any speech, which could hinder its performance under speech-heavy scenarios.
Lastly, we observe only little performance improvements when utilizing large data, most likely due to our model's small size, meaning that our future work would aim to improve the depth and complexity of our teacher/student models to utilize available data better.

\appendices
\section{Soft vs. dynamic labels}
\label{sec:soft_vs_dynamic_appendix}

In this paper, we utilized our dynamic method as the default label method without adequately providing results for $\mathcal{V}_{1},\mathcal{A}_1$ target datasets. 
In \Cref{tab:dynamic_vs_hard_appendix} these missing results can be seen.
Even though dynamic labels do not always provide better performance (e.g., on the clean A test set), a significant difference in terms of FER and Event-F1 can be seen between B and C test sets.
It seems that dynamic labels are much less prone to overfitting and are more capable to robustly estimate sound-event boundaries, evident by a similar Event-F1 score in tests B and C.
All provided results on the C test set consistently obtain an Event-F1 score of over 50\%, while their soft-label counterparts consistently obtain 40\%.


\section*{Acknowledgment}

This work has been supported by National Natural Science Foundation of China (No.61901265), Shanghai Pujiang Program (No.19PJ1406300), State Key Laboratory of Media Convergence Production Technology and Systems Project (No.SKLMCPTS2020003) and Startup Fund for Youngman Research at SJTU (No.19X100040009). Experiments have been carried out on the PI supercomputer at Shanghai Jiao Tong University.

\ifCLASSOPTIONcaptionsoff
  \newpage
\fi



\bibliographystyle{IEEEtran}
\bibliography{paper,sw121}
\end{document}